\documentclass[%
 reprint,
superscriptaddress,
floatfix,
 amsmath,amssymb,
 aps,
prx,
longbibliography,
]{revtex4-2}

\UseRawInputEncoding

\usepackage{graphicx}
\usepackage{dcolumn}
\usepackage{bm}
\usepackage[colorlinks]{hyperref}

\usepackage{graphicx} 

\usepackage{amsfonts}
\usepackage{amsmath}
\usepackage[english]{babel}
\usepackage{float}

\newcommand{\p}{\partial}

\newcommand{\mbf}{\mathbf}
\newcommand{\T}{\text}

\newcommand{\lb}{\langle}
\newcommand{\rb}{\rangle}

\begin{document}

\author{Matthew J. Hurley}
\altaffiliation{Contributed equally to this work}
\affiliation{Department of Physics, Arizona State University, Tempe, AZ 85287, USA}
\affiliation{Present Address: Department of Physics, Stanford University, Stanford, CA 94305, USA}

\author{Christian P. N. Tanner}
\altaffiliation{Contributed equally to this work}
\affiliation{Department of Chemistry, University of California, Berkeley, CA 94720, USA}

\author{Joshua Portner}
\affiliation{Department of Chemistry, James Franck Institute, and Pritzker School of Molecular Engineering, University of Chicago, Chicago, IL 60637, USA}

\author{James K. Utterback}
\affiliation{Department of Chemistry, University of California, Berkeley, CA 94720, USA}
\affiliation{Present Address: Sorbonne Université, CNRS, Institut des NanoSciences de Paris, INSP, 75005 Paris, France}

\author{Igor Coropceanu}
\affiliation{Department of Chemistry, James Franck Institute, and Pritzker School of Molecular Engineering, University of Chicago, Chicago, IL 60637, USA}

\author{Garth J. Williams}
\affiliation{Brookhaven National Laboratory, NSLS-II, Upton, NY 11973, USA}

\author{Avishek Das}
\affiliation{Department of Chemistry, University of California, Berkeley, CA 94720, USA}
\affiliation{Present Address: AMOLF, Science Park 104, 1098 XG, Amsterdam, The Netherlands}

\author{Andrei Fluerasu}
\affiliation{Brookhaven National Laboratory, NSLS-II, Upton, NY 11973, USA}

\author{Yanwen Sun}
\affiliation{Linac Coherent Light Source, SLAC National Accelerator Laboratory, Menlo Park, CA 94025, USA}

\author{Sanghoon Song}
\affiliation{Linac Coherent Light Source, SLAC National Accelerator Laboratory, Menlo Park, CA 94025, USA}

\author{Leo M. Hamerlynck}
\affiliation{Department of Chemistry, University of California, Berkeley, CA 94720, USA}

\author{Alexander H. Miller}
\affiliation{Department of Physics, Arizona State University, Tempe, AZ 85287, USA}

\author{Priyadarshini Bhattacharyya}
\affiliation{School for Engineering of Matter, Transport and Energy, Arizona State University, Tempe, AZ 85287, USA}


\author{Dmitri V. Talapin}
\affiliation{Department of Chemistry, James Franck Institute, and Pritzker School of Molecular Engineering, University of Chicago, Chicago, IL 60637, USA}
\affiliation{Center for Nanoscale Materials, Argonne National Laboratory, Argonne, IL 60517, USA}

\author{Naomi S. Ginsberg}
\altaffiliation{Co-corresponding author}
\email{nsginsberg@berkeley.edu}
\affiliation{Department of Chemistry, University of California, Berkeley, CA 94720, USA}
\affiliation{Department of Physics, University of California, Berkeley, CA 94720, USA}
\affiliation{Molecular Biophysics and Integrated Bioimaging Division, Lawrence Berkeley National Laboratory, Berkeley, CA 94720, USA}
\affiliation{Materials Sciences and Chemical Sciences Division, Lawrence Berkeley National Laboratory, Berkeley, CA 94720, USA}
\affiliation{Kavli Energy NanoSciences Institute, University of California, Berkeley, CA 94720, USA}
\affiliation{STROBE, NSF Science \& Technology Center, Berkeley, CA 94720, USA}

\author{Samuel W. Teitelbaum}
\altaffiliation{Co-corresponding author}
\email{SamuelT@asu.edu}
\affiliation{Department of Physics, Arizona State University, Tempe, AZ 85287, USA}

\title{In situ coherent X-ray scattering reveals polycrystalline structure and discrete annealing events in strongly-coupled nanocrystal superlattices
}


\begin{abstract}
Solution-phase bottom up self-assembly of nanocrystals into superstructures such as ordered superlattices is an attractive strategy to generate functional materials of increasing complexity, including very recent advances that incorporate strong interparticle electronic coupling. While the self-assembly kinetics in these systems have been elucidated and related to the product characteristics, the weak interparticle bonding interactions suggest the superstructures formed could continue to order within the solution long after the primary nucleation and growth have occurred, even though the mechanism of annealing remains to be elucidated. Here, we use a combination of Bragg coherent diffractive imaging and X-ray photon correlation spectroscopy to create real-space maps of supercrystalline order along with a real-time view of the strain fluctuations in aging strongly coupled nanocrystal superlattices while they remain suspended and immobilized in solution. By combining the results, we deduce that the self-assembled superstructures are polycrystalline, initially comprising multiple nucleation sites, and that shear avalanches at grain boundaries continue to increase crystallinity long after growth has substantially slowed. This multimodal approach should be generalizable to characterize a breadth of materials \textit{in situ} in their native chemical environments, thus extending the reach of high-resolution coherent X-ray characterization to the benefit of a much wider range of physical systems.
\end{abstract}


\keywords{
coherent X-ray scattering, coherent diffractive imaging, X-ray photon correlation spectroscopy, nanocrystal self-assembly, coarsening, \textit{in situ} measurement }

\maketitle
\section{Introduction}

Recently, a new class of bottom-up self-assembled nanomaterials has been developed that achieves a combination of long-range order and electrical conductivity by leveraging electrostatic, rather than steric, colloidal stabilization of individual nanocrystal (NC) building blocks \cite{coropceanu_self-assembly_2022,boles_self-assembly_2016,murray_self-organization_1995,shevchenko_structural_2006,smith_self-assembled_2009,bian_shape-anisotropy_2011,wang_colloids_2012,fan_self-assembled_2010,wang_self-assembled_2012,murray_synthesis_2000,santos_macroscopic_2021,van_blaaderen_template-directed_1997}. Electrostatic colloidal stabilization is obtained by binding compact, inorganic multivalent metallochalcogen ligands to the NC surface \cite{coropceanu_self-assembly_2022}. We previously studied the solution-phase assembly kinetics of such NCs \textit{in situ} with small-angle X-ray scattering (SAXS) and found fast (sub-ms) condensation of NCs into a superlattice (SL), a crystal of NCs, upon quenching via increasing the solution ionic strength \cite{tanner_situ_2024}. \textit{In situ} SAXS also shows that SL annealing progresses in solution at a small power law rate, but the nature of the coarsening events remains to be determined. Furthermore, the size and order of the resulting SL structures are seen to depend on the solution ionic strength, or quench depth, but the shape of the grains and nature of the observed heterogeneities during the annealing process are unknown, especially given the unusual interparticle interactions at play in this system \cite{coropceanu_self-assembly_2022}.

To assess the structure and heterogeneity of SLs during annealing and the nature of the annealing process itself requires direct space visualization of a three-dimensional (3D) object in solution at the nanoscale and characterization of similarly small fluctuations in the structure as a function of time. Coherent hard X-ray scattering \cite{veen_coherent_2005} techniques such as coherent diffractive imaging (CDI) \cite{fienup_reconstruction_1978,hayes_reconstruction_1982,miao_phase_1998,miao_extending_1999,gulden_coherent_2010,takahashi_three-dimensional_2010,pfeifer_three-dimensional_2006,williams_three-dimensional_2003,robinson_reconstruction_2001}, X-ray photon correlation spectroscopy (XPCS) \cite{dierker_x-ray_1995,thurn-albrecht_photon_1996,borsali_x-ray_2008,shpyrko_x-ray_2014,cao_effect_2020,robert_dynamics_2008,orsi_dynamics_2012,sutton_review_2008,sandy_hard_2018}, or a combination of the two \cite{takazawa_coupling_2023} in principle allow such characterizations. In this particular case, however, they also require a high degree of translational and rotational immobilization of individual condensed structures in a focused X-ray beam in solution for observation over at least tens of minutes.

Here, we capture a snapshot of an evolving SL formed from electrostatically stabilized NCs and also observe its annealing events to elucidate their nature. To achieve stability in solution for Bragg diffraction and CDI reconstruction, we developed a strategy to immobilize SLs in capillaries filled with quartz wool. Performing Bragg CDI \cite{xiong_coherent_2014,ulvestad_avalanching_2015,cherukara_three-dimensional_2018,vicente_bragg_2021} on the \textit{fcc} (111) SL peak, we resolve a $\sim$600 nm core of apparent high electron density, surrounded by a $\sim$few-micron apparently lower electron density region. We attribute the lower densities to crystallites that are not well-aligned with the measured (111) peak. Moreover, in using XPCS on the same SL, we are able to both validate the sufficient degree of immobilization in solution and elucidate the nature of SL annealing. Combining these findings, we conclude that SLs in solution form via a fast nucleation and growth step resulting in structures with a high density of grain boundaries. Subsequent annealing of these grain boundaries via discrete shearing events enables further ordering of the SLs.

\section{Results}

\begin{figure*}
    \includegraphics[]{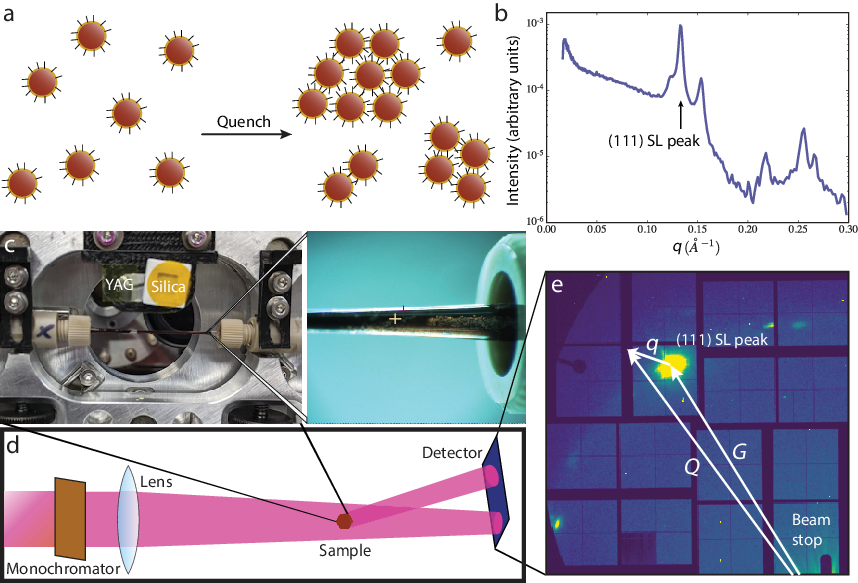}
    \caption{Experimental set-up and sample delivery. (a) Illustration of self-assembly of colloidal Au NCs (maroon dots) with Sn$_2$S$_6^{4-}$ ligands (black lines) into SLs (large clusters) upon quenching with an influx of excess salt. (b) Representative SAXS pattern from previous work showing colloid and SL coexistence, reproduced with permission \cite{tanner_situ_2024}. (c) Sample mount (left) for positioning of glass capillary (right) filled with quartz wool (silver mesh) in the X-ray beam. Yellow crosshair marks location of X-ray beam on capillary. (d) Schematic of the XCS instrument layout, not to scale. (e) Representative detector image collected during the experiment, integrated over many pulses. Here $\mbf Q$ is the scattering vector, $\mbf G$ the (111) reciprocal lattice vector, and $\mbf q=\mbf Q-\mbf G$ the reduced scattering vector.}
    \label{fig:Fig1_experimental_setup}
\end{figure*}

To perform the experiments, SLs of 4.6-nm diameter Au NCs capped with thiostannate (${\T{Sn}_2\T{S}_6}^{4-}$) ligands are prepared by adding 75 mM $(\T N_2\T H_5)_4\T{Sn}_2\T S_6$ salt to an electrostatically stabilized colloidal solution in hydrazine ($\T N_2\T H_4$), as described previously \cite{coropceanu_self-assembly_2022,tanner_situ_2024} (Fig. \ref{fig:Fig1_experimental_setup}(a)). A typical small-angle X-ray scattering (SAXS) intensity pattern, $I(Q)$, as a function of momentum transfer, $Q=|\mbf Q|$, of these samples measured previously is shown in Fig. \ref{fig:Fig1_experimental_setup}(b). Approximately 24 hours after the SLs are formed, they are injected into a thin-walled borosilicate capillary into which quartz wool has been inserted to immobilize them while maintaining their solvent environment to enable further evolution (Fig. \ref{fig:Fig1_experimental_setup}(c), Fig. \ref{fig:FigS1_microscopy}). The capillaries are securely sealed while still in an inert environment, and are mounted and moved into the center of rotation at the X-ray correlation spectroscopy (XCS) endstation of the Linac Coherent Light Source (LCLS) \cite{alonso-mori_x-ray_2015} (Fig. \ref{fig:Fig1_experimental_setup}(d)). The experiment was carried out at 11.8 keV, just below the Au absorption edge, using the Si(111) monochromator in a SAXS geometry with a sample-to-detector distance of 7.7 m, measuring $Q\sim0$ to 0.15 $\T\AA^{-1}$ in order to capture the (111) SL peak as shown in Fig. \ref{fig:Fig1_experimental_setup}(e). The X-ray beam, consisting of 50 fs pulses at 120 Hz, was focused with low divergence to $10 \times 10 \T{ }\mu$m$^2$ and attenuated to 5 $\cdot 10^9$ photon/s to preserve the integrity of the sample and its solution environment. To characterize aspects of the data we commonly refer below to the location in momentum space on the detector relative to the direct beam ($\mbf Q$) and also using a reduced scattering vector $\mbf q$, defined as $\mbf q = \mbf Q-\mbf G$, where $\mbf G$ is the (111) reciprocal lattice vector.

\begin{figure}[ht]
    \includegraphics[width=\columnwidth]{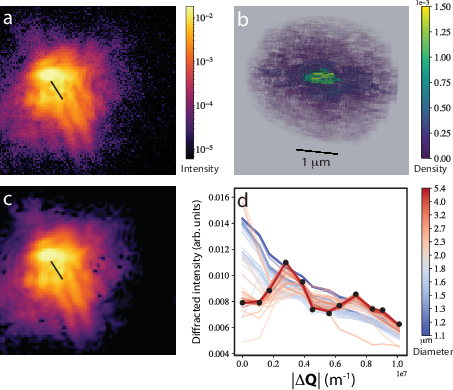}
    \caption{ Coherent diffractive imaging of a NC SL. (a) Detector image showing a 2D cross-section through the middle of the 3D intensity dataset (in arbitrary units), with a line cut of interest marked in black. (b) Surface visualizations of the reconstructed density, with scaled opacity. Yellow regions correspond to higher density and purple to lower density (in arbitrary units). Scale bar in black, indicating a distance of 1 $\mu$m. (c) Corresponding 2D cross-section of the reconstructed diffraction pattern, using the same color scale as (a). (d) Line cuts of the measured and reconstructed diffraction patterns. Black dots are the measured intensity as a function of distance in reciprocal space, $|\Delta \mbf Q|$, along the line cut in (a). Similarly, colored traces correspond to line cuts taken from reconstructed intensities, with color indicating the varied reconstruction support diameter.}
    \label{fig:Fig2_CDI}
\end{figure}

\textbf{Coherent diffractive imaging.} To elucidate the potentially heterogeneous structure and morphology of NC SLs, a full reciprocal space sampling of the SL (111) peak was collected. The sample was rotated over a 2$^{\circ}$ range in a total of 128 steps once a SL peak was identified on the detector. After suitable pre-processing, we then performed an error reduction/hybrid input-output phase retrieval algorithm \cite{siddharth_maddali_phaser_2020, xiong_coherent_2014} on the three-dimensional SL (111) peak (see \textbf{Methods}), with a representative detector image at a single angular step shown in \ref{fig:Fig2_CDI}(a). The initial object support (the region outside of which all density values are set to zero) was taken as a sphere of 1.5 $\mu$m  radius and enforced to be roughly spherical throughout the reconstruction process.

Fig. \ref{fig:Fig2_CDI}(b) presents the apparent three-dimensional real space electron density, replotted in an orthonormal coordinate basis (see Fig. \ref{fig:FigS2_sampling}). Each (cubic) voxel has a side length of 0.071 $\mu$m. The real space resolution is given by $2\pi$ divided by the dimensions of the sampled reciprocal space region and is then nearest-neighbor interpolated to obtain isotropic sampling. The corresponding reconstructed diffraction pattern, obtained via Fourier transform of the electron density, is shown in \ref{fig:Fig2_CDI}(c).
Comparison between \ref{fig:Fig2_CDI}(a) and \ref{fig:Fig2_CDI}(c) shows good agreement between the measured and reconstructed diffraction patterns, especially in brighter regions where the signal-to-noise ratio in the data is higher. In particular, the reconstructed pattern preserves the fringe spacing and overall contours of the data. It also features singularities (dark spots) in the already low-intensity regions of the diffraction pattern that are not clearly visible in the measured detector image. These characteristics persist across reconstructions with different random initial starting points. The apparent density of the reconstruction in \ref{fig:Fig2_CDI}(b) features a higher-density central region surrounded by a few-micron shell over which the density tapers off by more than an order of magnitude (Fig. \ref{fig:FigS3_135RSL}). In an effort to understand whether the lower-magnitude outer regions seen in pale blue in \ref{fig:Fig2_CDI}(b) are physical or simply a reconstruction artifact, we perform a set of reconstructions with various support sizes in which the support is unaltered during the reconstruction algorithm. \ref{fig:Fig2_CDI}(d) compares a line cut of the data (black circles) to line cuts of the Fourier transform of the reconstructed density as a function of support diameter. The reconstructed reciprocal space intensities reproduce the features of the data well only for sufficiently large supports that extend well beyond the bounds of the higher-magnitude central region of the density. This analysis therefore suggests that there is indeed a physical basis for the lower-magnitude region featured in \ref{fig:Fig2_CDI}(b). Furthermore, we obtain similar results from data collected on a different SL in the same capillary (see Figs. \ref{fig:FigS4_116cdi}, \ref{fig:FigS5_116RSL} and accompanying text). We defer proposing origins of this apparent lower-magnitude density region to the \textbf{Discussion} in order to first present the XPCS results for this same SL structure.

\begin{figure}
    \includegraphics[width=\columnwidth]{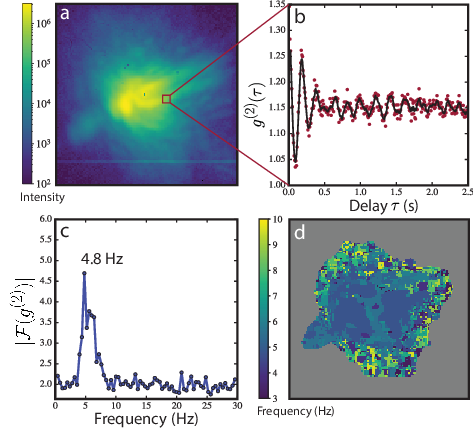}
    \caption{Short-time oscillatory behavior in the SL (111) intensity autocorrelation function. (a) Cropped detector image featuring the (111) Bragg peak. (b) Intensity autocorrelation $g^{(2)}(\tau)$ calculated over the red ROI marked in (a), five-point rolling average in black. (c) Fourier transform amplitude of the signal in (b). (d) Frequency map of the detector image in (a), showing the strongest oscillatory component in the autocorrelation spectral density at each pixel.}
    \label{fig:Fig3_XPCS_oscillations}
\end{figure}

\textbf{X-ray photon correlation spectroscopy.} XPCS serves two main purposes in our study; we use it first to assess the extent of SL immobilization over the course of the measurements and, second, to characterize the nature of SL annealing. To assess the extent of SL immobilization by the quartz wool via XPCS, detector images from individual X-ray pulses were collected over a period of 140 seconds. \ref{fig:Fig3_XPCS_oscillations}(a) displays a cropped detector image of a SL (111) peak averaged over the full dataset. Following pixel calibration \cite{van_driel_epix10k_2020} (see \textbf{Methods}), the temporal autocorrelation in Fig. \ref{fig:Fig3_XPCS_oscillations}(b) was obtained from the red boxed 5 $\times$ 5 pixel ROI indicated in Fig. \ref{fig:Fig3_XPCS_oscillations}(a) using the multi-tau correlation algorithm, which averages shots together prior to correlation for long delays \cite{abeykoon_software_2016,lumma_area_2000}. We return to an analysis of the longer-delay behavior of such temporal autocorrelations below to consider the nature of the SL annealing events. At short delays, the autocorrelation in Fig. \ref{fig:Fig3_XPCS_oscillations}(b) features an oscillation, which appears as a peak in the Fourier spectrum (Fig. \ref{fig:Fig3_XPCS_oscillations}(c)) at approximately 5 Hz. Calculating an autocorrelation over each 5$\times$5 pixel region on the detector and identifying its strongest-peaked frequency between 3 Hz and 10 Hz yields a frequency map of the Bragg peak in Fig. \ref{fig:Fig3_XPCS_oscillations}(d). A lower signal-to-noise ratio away from the peak makes oscillatory signals hard to resolve, and the peak frequencies in those regions vary widely. The central portion of the peak, however, consists of more homogeneous, and therefore more reliable, values with an average of $\sim5 \pm 1$ Hz.

\begin{figure*}
    \includegraphics[width=\textwidth]{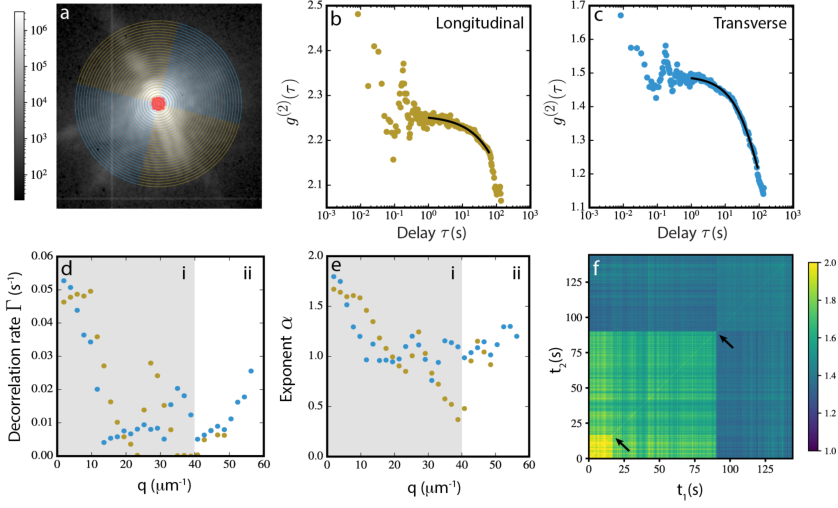}
    \caption{ Observation of SL annealing via XPCS. (a) Cropped detector image featuring the (111) Bragg peak. Overlaid in blue (gold) are annular ROIs along the transverse (longitudinal) direction. (b, c) Representative intensity autocorrelations for a longitudinal and transverse ROI (gold, blue points respectively), each fit to a stretched exponential (black curve). (d) The decorrelation rate, $\Gamma$, as a function of $q$ for the longitudinal (gold) and transverse (blue) directions. We divide the dispersion into two regimes, (i) (gray) and (ii) (white), based on the inverse length scale $q$. (e) The stretching exponent $\alpha$ as a function of $q$ for each of the longitudinal (gold) and transverse (blue) directions, with the same marked regimes (i) and (ii). (f) The two-time correlation function of the data shown in (c), featuring intensity jumps (marked by arrows).}
    \label{fig:Fig4_XPCS_avalanches}
\end{figure*}

To characterize aspects of SL annealing we performed additional analysis on the dataset shown in Fig. \ref{fig:Fig3_XPCS_oscillations}. To elucidate the scale-dependent dynamics within the SL, we calculate correlation functions as a function of distance from the Bragg peak, as shown in Fig. \ref{fig:Fig4_XPCS_avalanches}(a). We divide the detector ROI containing the peak into annular regions bounded by contours of constant $q = |\mbf Q-\mbf G|$, and we further subdivide each region along the longitudinal and transverse components of $q$, marked in gold and blue, respectively. In this SAXS geometry, the longitudinal direction is parallel to $\mbf G$, and the transverse direction is tangent to the (111) diffraction ring. For each annular region, we fit the corresponding autocorrelation function, $g^{(2)}(q,\tau)$, to a compressed/stretched exponential of the form $g^{(2)}(q,\tau) = k(q) + \beta(q)\exp[-2(\Gamma(q)\tau)^{\alpha(q)}]$, where $k(q)$ is a baseline, $\beta(q)$ is the speckle contrast, $\Gamma(q)$ is the decorrelation rate, and $\alpha(q)$ is a stretching exponent. Fig. \ref{fig:Fig4_XPCS_avalanches}(b)-(c) shows example autocorrelations and fits for respective longitudinal and transverse directions at $q=18 \T{ }\mu\T{m}^{-1}$ (see also Figs. \ref{fig:FigS6_radialfits}, \ref{fig:FigS7_radialfits}, \ref{fig:FigS8_singlecenterdispersion} for data and fits at additional $q$ values). Such stretched/compressed exponentials often appear in systems with heterogeneous dynamics, where the resulting distribution of time scales leads to distortion of the simple exponential decay or Gaussian correlation forms \cite{madsen_beyond_2010}. Note that a speckle contrast higher than 1 arises when there are large variations in intensity (e.g., diffraction fringes) within the ROI. Because this analysis depends on the choice of center for the annular regions of interest and there is ambiguity in how this choice should be made, we repeat the analysis for many different choices of center, using pixels within a circular region (indicated in red in Fig. \ref{fig:Fig4_XPCS_avalanches}(a)) with a 5 pixel radius around the intensity center of mass, and averaging the resultant fit parameters together. Fig. \ref{fig:Fig4_XPCS_avalanches}(d) shows the decorrelation rate $\Gamma$ as a function of $q$ for each of the radial and transverse directions, respectively. The corresponding stretching exponents, $\alpha$, are shown in Fig. \ref{fig:Fig4_XPCS_avalanches}(e). The behavior of $\Gamma$ at low $q$ (region i, in gray) is complex and non-monotonic. The dispersion contrasts with that of, for instance, a simple, freely-diffusing colloidal system, where $\Gamma\sim q^2$ \cite{borsali_x-ray_2008}. Unlike a diffusive system, which has no characteristic length scale, however, our system has a clear length scale of relevance, namely the size of the SL. For $q$ values around this value we would not anticipate a simple dispersion. But at sufficiently small length scales we should expect to recover monotonicity, which we indeed observe at high $q$ (region ii, white). Here the identification of the two regions is merely to highlight the point between these regimes at which such conventional, monotonic behavior manifests in the correlation functions. In the transverse direction, the decorrelation rate is on the order of $10^{-2}$ Hz and increases with increasing $q$, while in the longitudinal direction we do not resolve such dispersive behavior (and cannot reliably extract fits at the highest $q$ values). This trend in decorrelation rate in the transverse direction, and the lack thereof in the longitudinal direction, persist even if one fits to a fixed exponent of $\alpha=1$ instead (Fig. \ref{fig:FigS9_fixedfits}).  In order to investigate the source of this wavevector-dependent time scale, we calculate a two-time correlation function in Fig. \ref{fig:Fig4_XPCS_avalanches}(f) using the ring-shaped region of Fig. \ref{fig:Fig4_XPCS_avalanches}(a) around $q=18\T{ }\mu\T{m}^{-1}.$ Two-time analysis enables us to observe from when in the data acquisition period the time scale arises and whether it is irreversible \cite{plumley_speckle_2020, widera_non-ergodic_2015,sanborn_direct_2011}. The raw data were first binned down to 0.1 s intervals to more readily visualize larger jumps in intensity. We observe distinct square-like patterns in the two-time correlation, corresponding to sudden shifts in the diffraction intensity at 17 s and 90 s.

\section{Discussion}
Having described the CDI and XPCS results, we now discuss possible interpretations and their implications for elucidating the self-assembled NCs’ products and their annealing. First, we comment on the quartz wool matrix immobilization strategy for coherent scattering characterization of individual structures in the solution phase and in particular on the sufficiency of immobilization of the structures that we interrogated. Second, we discuss the hypothesis that the discrete time scales observed in the XPCS correspond to sudden shear-like avalanches. Third, we propose and weigh various interpretations of the direct space reconstruction obtained from CDI and in particular the origin of the apparent low-density shell surrounding the higher density core. Finally, we discuss the implications from assembling our CDI and XPCS findings self-consistently by inferring the mechanism by which the observed structures are nucleated, grown, and further evolved.

This work extends the reach of coherent diffractive imaging by probing a solid, immobilized structure in its native liquid environment \cite{grote_imaging_2022}. To be able to extract meaningful conclusions about SL structure and evolution from our diffraction data via CDI and XPCS analysis techniques, the SLs must be sufficiently immobilized in the X-ray beam. The low-$q$ behavior of the SL Bragg peak is governed by long-range order in the SL; thus the oscillations that we observe in the autocorrelation functions in \ref{fig:Fig3_XPCS_oscillations}, which feature prominently at low $q$, are likely produced by a bulk motion of the SL, such as small librations of the SL in solution. Perturbed by solvated particles or external vibrations, the SLs sway periodically on the quartz wool, producing intensity fluctuations at a rate much slower than the X-ray pulse repetition rate of 120 Hz. While the SLs are therefore not perfectly immobilized in the capillary, these perturbations are quite small (see \textbf{Appendix F} and Figs. \ref{fig:FigS10_rockingcurves}, \ref{fig:FigS11_inplanelinecuts}), and we are nevertheless able to resolve sharp reciprocal space features in our dataset, an encouraging sign that the sample is sufficiently immobilized for the purposes of our XPCS and CDI analyses.

We now turn to a discussion of the long-time behavior of the SLs, as revealed by our XPCS results. We first observe that the dispersive increase (Fig. \ref{fig:Fig4_XPCS_avalanches}(d)) at high $q$ in the transverse fit parameters extracted from the XPCS data reflects conventional dynamical behavior in which scattering decorrelates more quickly at higher wavevector ($\mbf q$) \cite{borsali_x-ray_2008}. Moreover, wavevectors in the transverse direction are sensitive to shear-like motions of the SL, analogous to the sensitivity to transverse acoustic phonons near Bragg peaks \cite{warren_x-ray_1990,baron_introduction_2015}. Therefore, since we observe this dispersion along the transverse direction but not along the longitudinal direction, we interpret the time scale as that of shear-like fluctuations of the SL. We see, furthermore, that this one-time decorrelation is accompanied by square-like patterns in the two-time correlation (Fig. \ref{fig:Fig4_XPCS_avalanches}(f)), a phenomenon that we also observe in an additional experimental dataset (Fig. \ref{fig:FigS12_ttcf93}). In a third experimental dataset (Figs. \ref{fig:FigS13_XPCS_Oscillations_115}, \ref{fig:FigS14_115g2s}), we do not observe any decay in the autocorrelation functions at long delay, and the corresponding two-time correlation function (Fig. \ref{fig:FigS15_115ttcf}) is highly uniform and featureless, displaying no such rapid discontinuities as those seen in Fig. \ref{fig:Fig4_XPCS_avalanches}(f). Since the absence or presence of a one-time decorrelation rate is thus accompanied by the absence or presence of these discontinuities in the two-time correlation, the decorrelation is likely dominated by sudden changes in the diffraction pattern instead of small fluctuations therein. While we cannot rule out the possibility that these sudden changes in the diffraction pattern are merely due to bulk translations or rotations of the SL in solution, the increasing dispersion in decorrelation rate at high $q$ is indicative of internal SL dynamics.  Furthermore, from prior ensemble SAXS measurements on NC SLs \cite{tanner_situ_2024}, we expect SLs to continue to evolve (via growth and defect annealing) even many hours after they are initially nucleated, provided their solution phase environment is maintained. Taken together, these features in the data suggest that the high-$q$ fluctuations in the SL Bragg peak are driven by sudden shear-like ``avalanches'' \cite{argon_strain_2013,sethna_deformation_2017} of NCs rather than a continuous annealing process. Such events are roughly analogous to the microstructural avalanches previously observed in martensitic metallic alloys \cite{sanborn_direct_2011,widera_non-ergodic_2015}. In our previous \textit{in situ} SAXS experiments of SL formation, SLs were shown to evolve at different rates depending on the quench depth at which they were formed. At shallow quenches, highly ordered SLs formed but did not evolve much after their initial formation. In contrast, at slightly deeper quenches, more disordered SLs form initially but they continue to evolve over the course of the several hour observation period \cite{tanner_situ_2024}. This observation is consistent with a picture in which annealing proceeds via such shear-like avalanches, in which these strain-relieving events are more likely to occur in regions of high disorder \cite{porta_intermittent_2018,romero_thermo-magnetic_2021,ricci_intermittent_2020}. Moreover, previous ensemble measurements also find that the SL Bragg peak width evolves according to a power law over time, a relationship we would expect to arise from the occurrence of avalanches within many SLs within an ensemble throughout a bulk measurement window \cite{fisher_collective_1998}.

Returning to the real space reconstruction obtained from CDI presented in Fig. \ref{fig:Fig2_CDI}, we consider interpretations of the decrease in electron density that arises moving outward from the center of the SL. One possibility is that the reconstruction algorithm has failed, assigning low non-zero values to regions of real space that extend beyond the boundary of the object. Another possibility is that the diffraction data is blurred due to the SL librations discussed above, thus yielding artifacts on large length scales in the reconstruction. A further alternative is that inhomogeneities in the beam path (due to the presence of solvent, quartz wool, and capillary glass) induce an effective decoherence in the beam and distort the diffraction signal. There are several reasons, however, to believe that the low-density regions are not an experimental artifact. First, the analysis of size-dependence presented in \ref{fig:Fig2_CDI}(d) suggests that there is diffracting material over a span of 2-3 $\mu$m, extending beyond the high density region of the reconstruction.  Additionally, we observe similar such low-density regions in the second CDI dataset that we analyzed (Fig. \ref{fig:FigS5_116RSL}). For this latter dataset the reconstructed object, including the low-density region, is smaller, in accordance with a larger fringe spacing. If beam contamination were the driving force behind a reconstruction artifact, we would expect to see its impact appear on similar length scales across all datasets. Furthermore, we can estimate an upper bound for the magnitude of phase distortion in the scattered beam. For quartz wool fibers of 9-30 $\mu$m, supposing that the part of the beam passes through at worst the edge of one fiber, we would expect a relative phase shift of around $\Delta \phi < \pi/6$. This places an upper bound on the phase distortion due to the quartz wool, and is small enough to preserve a good CDI signal (see \textbf{Appendix C} for quantitative estimates). One physical interpretation of these apparent lower electron densities is that they are produced by regions of greater disorder in the SL. By analyzing and comparing hypothetical ordered and disordered SL structures, we attempt to assess the impact of at least one kind of defect on such a Bragg CDI reconstruction.

\begin{figure*}
    \includegraphics[width=\textwidth]{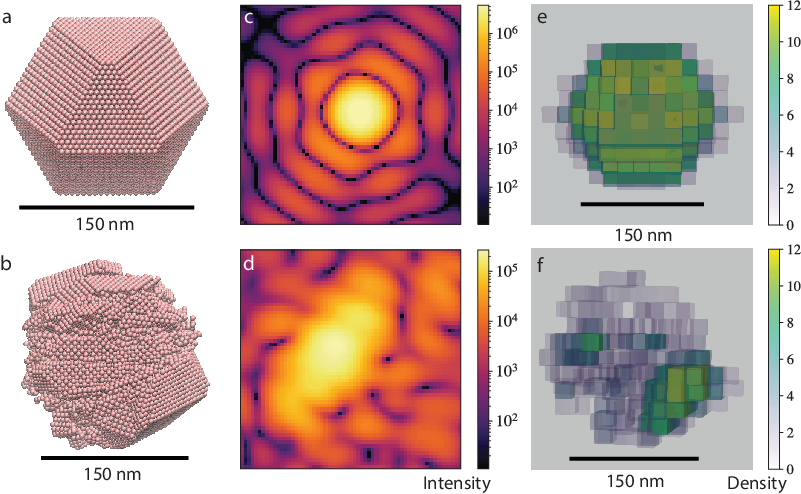}
    \caption{Comparative analysis of perfect and defective SLs. (a) Real space depiction of a perfectly ordered single crystal \textit{fcc} lattice. (b) Real space structure of a polycrystalline SL structure. (c, d) 2D intensity patterns (arbitrary units) obtained via Fourier transform of the structures in (a) and (b), respectively, featuring the (111) peak. (e, f) 3D visualization of the inverse Fourier transform magnitude (arbitrary units) of the (111) diffraction peak corresponding to each of (a) and (b), respectively. All scale bars (black) denote a real space distance of 150 nm.}
    \label{fig:Fig5_simulations}
\end{figure*}

Toward this end, we introduce and then thermally relax planar defects (grain boundaries) into a simulated \textit{fcc} lattice, representing ordered and disordered SL structures that are $\sim$150 nm in scale. In order to reduce computational load we simulate smaller structures than in experiment. Doing so does not preclude qualitative comparison with experiment, as the grain boundaries are topological objects, and their effect on the CDI reconstruction is independent of SL size. Fig. \ref{fig:Fig5_simulations}(a) and \ref{fig:Fig5_simulations}(b) show the real space atomic positions of the original (single crystal) and polycrystalline lattices, respectively. The polycrystal’s scattering pattern is not peaked at a single reciprocal lattice vector; we choose the most intense (111) Bragg peak and plot a two-dimensional cross-section thereof in \ref{fig:Fig5_simulations}(d), with the (111) peak of the single crystal lattice shown in \ref{fig:Fig5_simulations}(c). Note that the detector image in \ref{fig:Fig2_CDI}(a), with its lack of centrosymmetry and its irregular fringes, more closely resembles the diffraction pattern of the polycrystalline lattice than that of the single crystal. We next take the inverse Fourier transform of a three-dimensional rectangular subregion about each diffraction peak in \ref{fig:Fig5_simulations}(c)-(d), using the computed scattering phases, to obtain the real space distributions shown in \ref{fig:Fig5_simulations}(e)-(f). While the real space density obtained from the perfect crystal is nearly constant across the object, that of the polycrystal features large variations. In particular, in domains of the polycrystal that are well-aligned with the chosen (111) peak, the density attains a value as large as that of the single crystal, whereas in misaligned regions the density drops by an order of magnitude or more. That is, in performing phase retrieval on the (111) peak alone, rather than on the full object’s Fourier transform, only those components of the object that contribute to the (111) peak appear in the reconstruction, and thus the reconstructed ``electron density'' is a measure of the density of planes aligned with the (111) direction at each point in the object rather than of the raw NC packing fraction. This observation also suggests that the lower-density regions are not due to residual scattering from the capillary, solvent, or quartz wool.

The combination of CDI and XPCS analysis together has proven to be a valuable means to obtain a coherent description of a self-assembled SL structure and its evolution in solution that could not otherwise have been realized. We previously found via \textit{in situ} SAXS that initial SL nucleation takes place on a rapid (sub-second) time scale, and then growth and annealing progress over the course of minutes and hours. Taking all of our findings together, one interpretation of the apparent density profile we observe from CDI is that the SL crystallinity decays away from its center, with regions of greater disorder towards the surface. While we cannot rule out this possibility based on the reconstructions alone, we would not expect such a disordered shell to be so large in proportion to the crystalline core. Although other phases such as gels and liquids have been implicated as products of NC self-assembly \cite{coropceanu_self-assembly_2022}, we found no evidence in this work and in previous work with the same exact metal NC system \cite{tanner_situ_2024} of these non-crystalline condensed phases. In addition, real space optical and electron microscopy images of this system following a quench show large, micron sized polycrystalline structures (Fig. \ref{fig:FigS1_microscopy}). Therefore a more likely explanation is that the SL (111) peak analyzed in Fig. \ref{fig:Fig2_CDI} is from a single crystal grain of a larger cluster of aggregated SLs, each of which nucleated separately; that is, the reconstruction ``core'' is simply one large, optimally oriented domain among many. Our analysis of hypothetical polycrystals in Fig. {\ref{fig:Fig5_simulations} demonstrates that in performing CDI on a polycrystalline structure whose other grains do not meet the Bragg condition, we would expect to see regions of apparent low density in the reconstructed object. This polycrystalline picture also accords well with the account of sub-lattice dynamics indicated by correlation analysis and with the structures observed in electron microscopy in Fig. \ref{fig:FigS1_microscopy}(c)-(d). In particular, the presence of defects such as slip planes or grain boundaries between neighboring crystallites can lead to a build-up of strain in the SL. Once this strain becomes sufficiently large, annealing should occur via discrete, large-scale shear motions. Based on the data analyzed here and in previous work, we thus propose that Au NC SLs quickly nucleate and then coalesce into polycrystalline structures, with shear annealing occurring within and/or between crystallites via strain-relieving avalanches.

\section{Conclusion}

In this work, we demonstrate Bragg-CDI and XPCS on the (111) SL peak of self-assembled strongly-coupled Au NCs in the solution phase. By immobilizing the SLs in quartz wool, we are able to observe annealing-like events and to obtain SL Bragg-CDI reconstructions of their real-space density. Using these techniques we uncover a picture of SL structure and evolution that accords well with previous ensemble SAXS measurements and microscopy, in which SLs agglomerate into polycrystalline structures after initial nucleation, and become further ordered over time via shear annealing avalanches.

Only recently have such strongly-coupled SLs been synthesized from electrostatically stabilized NCs. The large parameter space at play in SL formation, if fully understood, offers unique control over the coupling between NCs and the resultant optoelectronic properties of the material, enabling access to an electronic regime beyond that of individual quantum dots or bulk semiconductors \cite{lazarenkova_miniband_2001,kagan_building_2016}. The nature of the constituent NCs (e.g., metallic or semiconducting) dramatically alters the interparticle interactions, mechanism of self-assembly, and nature of the defects and annealing mechanisms within a SL. Bragg-CDI offers an incisive, \textit{in situ}, real-space tool to interrogate this diverse and unexplored chemistry, which takes place on unusual length scales much larger than those of conventional materials. A deeper understanding of these facets of self-assembly will facilitate the adoption of such strongly-coupled SLs as tunable, functional materials \cite{talapin_prospects_2010, vanmaekelbergh_self-assembly_2011,boles_self-assembly_2016}.

More broadly, this experiment demonstrates the successful performance of coherent diffractive imaging on a solid structure in solution. The sensitive reaction conditions at play here and in other chemical systems pose a challenge for most X-ray visualization methods, rendering it difficult to obtain detailed structural information in the structure's native environment. By demonstrating the feasibility of CDI for \textit{in situ} liquid phase measurements, this work opens the door for material imaging experiments of increasing complexity and sophistication. Furthermore, while XPCS is conventionally used for weakly-interacting systems, here we show its utility in probing supercrystalline structures on their fundamental length and time scales. Only via the combination of both XPCS and CDI together was it possible to uncover a complete picture of long-time annealing and reconstruction processes; this multimodal approach is thus a powerful means of elucidating the structural evolution of material systems at sufficiently low time scales.


\section{Acknowledgments}

We thank Y. Cao for thoughtful discussions and advice. This work was supported by the Office of Basic Energy Sciences (BES), US Department of Energy (DOE) (award no. DE-SC0019375). Use of the LCLS is supported by the U.S.
Department of Energy, Office of Science, Office of Basic
Energy Sciences under contract no. DE-AC02-76SF00515. This research also used resources of the National Synchrotron Light Source II, a U.S. Department of Energy (DOE) Office of Science User Facility operated for the DOE Office of Science by Brookhaven National Laboratory under Contract No. DE-SC0012704. C.P.N.T. was supported by the NSF (Graduate Research Fellowship no. DGE1106400). J.K.U. was supported by an Arnold O. Beckman Postdoctoral Fellowship in Chemical Sciences from the Arnold and Mabel Beckman Foundation. L.M.H. acknowledges a National Defense Science and Engineering Graduate Fellowship. A.D. was supported by a Philomathia Graduate Student Fellowship from the Kavli Nanoscience Institute at UC Berkeley.
N.S.G. was supported by a David and Lucile Packard Foundation Fellowship for Science and Engineering and a Camille and Henry Dreyfus Teacher-Scholar Award.

\section*{Appendix A: Methods}
\subsection{SL self-assembly.} To assemble the SLs we added $(\T N_2\T H_5)_4\T{Sn}_2\T S_6$ to a solution of colloidal gold NCs (of 4.6 nm diameter) in hydrazine in a vial in a nitrogen-filled glovebox. The solution after quenching had an Au NC concentration of 50 mg/mL and a $(\T N_2\T H_5)_4\T{Sn}_2\T S_6$ concentration of 75 mM. Self-assembly progressed in the vial for 24 hours, after which the solution was transferred to a borosilicate glass capillary (0.9 mm outer diameter and 0.860 mm inner diameter) and sealed with inert plugs. To immobilize the SLs in their native solution, the capillary was filled with quartz wool (9-30 $\mu$m in diameter) prior to the addition of the SL solution. The sealed capillaries were brought to the X-ray Correlation Spectroscopy (XCS) endstation at the Linac Coherent Light Source (LCLS) for measurement.

\subsection{Time-resolved X-ray scattering.} Coherent X-ray scattering data were collected at the XCS endstation of the Linac Coherent Light Source \cite{alonso-mori_x-ray_2015,vartanyants_coherence_2011}. The experiment was carried out at 11.8 keV, just below the Au absorption edge, using the Si(111) monochromator in a SAXS geometry with a sample-to-detector distance of 7.7 m, measuring $Q\sim0$ to 0.15 $\T\AA^{-1}$. The X-ray beam, consisting of 50 fs pulses at 120 Hz, was focused via a compound refractive lens to $10 \times 10 \T{ }\mu$m$^2$ and attenuated to 5 $\cdot 10^9$ photon/s. A double-crystal Si(111) monochromator was used to produce a monochromaticity of $\Delta E/E = 1.4\cdot 10^{-4}$. CDI data collection consisted of 128 angular steps over a two degree rocking range. We collected about two seconds of data collection at each angular step. All data were subject to a detector calibration pipeline consisting of gain, pedestal, threshold, and common mode corrections \cite{van_driel_epix10k_2020}. All shots at a given rocking angle are summed together to obtain a single image for each angular step; we then normalize each summed image by the sum of the corresponding intensity and position monitor (IPM) values, which record the intensity of the incident beam of each shot.

\subsection{CDI analysis.} The Python package Phaser was used to carry out error reduction (ER), hybrid input-output (HIO), and shrinkwrap algorithms \cite{siddharth_maddali_phaser_2020}. The reconstruction in \ref{fig:Fig2_CDI}(b) was produced using an initial support that is a sphere of radius 1.5 $\mu$m. We used the following algorithm to obtain reconstructions: 3x[500 iterations ER, 200 iterations HIO], 500 iterations ER, 10x[Shrinkwrap/intersect, 5x[500 iterations ER, 200 iterations HIO], 500 iterations ER], 10x[500 iterations ER, 200 iterations HIO], 2000 iterations ER (see \textbf{Appendix C} for details of the Shrinkwrap/intersect step). In visualizing the reconstruction in \ref{fig:Fig2_CDI}(b) we use nearest-neighbor interpolation to correct for the non-unitary sampling in real space. The sampling vectors are sufficiently close to orthogonal that we do not correct for non-orthogonality. The reconstructions in \ref{fig:Fig2_CDI}(d) were produced in a similar manner. For more details on the reconstructions, see \textbf{Appendix C}.

\subsection{XPCS analysis.} For XPCS data analysis, all one-time and two-time correlations were calculated using the Python package scikit-beam \cite{abeykoon_software_2016} using a symmetric normalization scheme. We divide the detector into ring-shaped regions of width 2.5 pixels, centered at a given point the peak, and further divide each ring into longitudinal and transverse sections. To compute the location of $\mbf Q=0$, we fit the (111) diffraction ring to a circle, and take the direct beam to be at the center. We calculate an autocorrelation for each ROI and fit to a stretched exponential form. We repeat this fitting process for each pixel within 5 pixels of the intensity center of mass of the time-integrated intensity pattern. We then remove outliers from the fit results (points at least 1.5 times the interquartile range larger/smaller than the upper/lower quartile) and average the remaining parameters together to get the final results.

\subsection{Computational methods.} For the simulated hypothetical structures in \ref{fig:Fig5_simulations}, we arrange 19619 atoms into an \textit{fcc} lattice. To produce the polycrystal, we randomly select 20 dividing planes that are either (111), (110), or (100) and, for each one, rotate one side of the crystal by a random angle, and shift the rotated domain a particle radius away perpendicular to the plane, to prevent particle overlap. We then minimize the energy of the resulting grain boundaries using a pairwise Lennard-Jones potential. From here, we use the NC positions to compute the structure factor and multiply by a hard-sphere form factor to obtain the complex scattering pattern. We perform a numerical inverse Fourier transform on a cubic region of interest containing the Bragg peak, the complex magnitude of which yields the density plots in Fig. \ref{fig:Fig5_simulations}(e) and \ref{fig:Fig5_simulations}(f).


\section*{Appendix B: Sample preparation and delivery} To assemble the SLs we added $(\T N_2\T H_5)_4\T{Sn}_2\T S_6$ to a solution of colloidal gold nanoparticles (of diameter 4.6 nm) in hydrazine in a vial in a glovebox. The solution after quenching had an Au NC concentration of 50 mg/mL and a $(\T N_2\T H_5)_4\T{Sn}_2\T S_6$ concentration of 75 mM. Self-assembly progressed in the vial for 24 hours, after which the solution was transferred to a glass capillary and sealed with paraffin wax and epoxy. The SLs in the sealed capillaries were then immediately transported to the beamline. To immobilize the SLs in their native solution, a borosilicate capillary with an outer diameter of 0.9 mm and a wall thickness of 20 um was filled with quartz wool of fiber thickness between 9 and 30 um, prior to the addition of the SL solution. Fig. \ref{fig:FigS1_microscopy} shows an image of a capillary and optical microscopy images of SLs immobilized in the quartz wool. Note that these images are representative and do not correspond exactly to the data collected and presented in the main text. Moreover, Fig. \ref{fig:FigS1_microscopy}c presents scanning electron microscopy images on a separate set of Au SLs, featuring aggregated polycrystalline structures. Fig. \ref{fig:FigS1_microscopy}d is a transmission electron microscope image of an Au SL, reproduced from \cite{coropceanu_self-assembly_2022}.

\begin{figure}[h!]
    \center
    \includegraphics[width=\columnwidth]{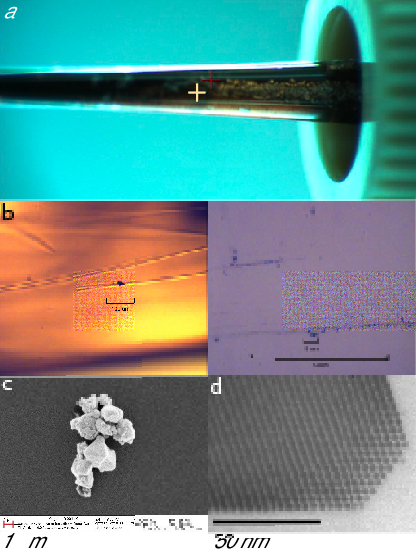}
    \caption{Microscopic images of Au SLs. (a) Glass capillary, without quartz wool, featuring SL clusters (dark specks). (b) Microscope images of quartz-filled capillary contents, featuring clumps of SL on quartz wool fibers. (c) Scanning electron microscopy image of SL clusters; 1 $\mu$m scale bar in red. (d) A transmission electron micrograph of the edge of a SL grain, where the SL is thin enough to resolve structure; 50 nm scale bar in black. Reprinted in part with permission from ref 1. Copyright 2022 The American Association for the Advancement of Science.}
    \label{fig:FigS1_microscopy}
\end{figure}

\newpage
\section*{Appendix C: Time- and momentum-resolved coherent scattering experiments} As an X-ray free electron laser, LCLS is one of the brightest X-ray sources in existence, offering X-ray flux at levels orders of magnitude larger than that of synchrotrons. The XCS instrument is used for time-resolved coherent scattering experiments. The double-crystal Si(111) monochromator produces a monochromaticity of $\Delta E/E = 1.4\cdot 10^{-4}.$ The detector sits on a long arm which can be angled to collect scattering data away from the forward direction. For this experiment, we use the ePix10k 2-megapixel detector, which has $100 \times 100\T{ }\mu$m pixels arranged in 16 modules, each $352 \times 384$ pixels \cite{van_driel_epix10k_2020}.

\section*{Appendix D: CDI methods and additional analysis}
\subsection*{Data collection and processing}
To collect CDI data, the sample is rotated (``rocked'') about an axis normal to the incident wavevector, with shots taken at a range of different values of the rocking angle. For a given rocking angle, the detector image captures a two-dimensional cross-section of the 3D Fourier transform (Bragg peak) in reciprocal space. By rocking through only a couple of degrees, we sample the entire Bragg peak in the form of a series of near-parallel two-dimensional detector images, yielding the three-dimensional reciprocal space data that is then subject to phase retrieval algorithms to obtain the real space SL structure. For this experiment, CDI data collection consisted of 128 angular steps (each an approximately 0.016 degree rotation) over a two degree rocking range. We collected 243 shots at 120 Hz at each angular step, or about two seconds of data collection per step, before moving to the next angle.

All data was subject to a detector calibration pipeline consisting of gain, pedestal, and common mode corrections \cite{alonso-mori_x-ray_2015}. Since the photon energy was 11.8 keV, we set all pixels with calibrated intensity values smaller than 7 keV to zero. We bin the data based on the rocking angle; i.e., all shots at a given rocking angle are summed together to obtain a single image for each angular step. We then normalize each binned image by the sum of the corresponding intensity and position monitor (IPM) values, which record the intensity of the incident beam of each shot. After binning, the entire detector image is cropped down to a region containing just the Bragg peak to be used for phase retrieval (a 114$\times$114 pixel region). Dead pixels and hot pixels were identified and interpolated using 2D cubic interpolation across each detector image; any negative values after interpolation were set to zero.

\newpage
\subsection*{Reciprocal space sampling calculation }
CDI data collection yields a set of intensity values sampled from a three-dimensional region of reciprocal space. In the general case of Bragg CDI the sampling is neither unitary (i.e., the sampling vectors have different lengths) nor orthogonal. In our experimental SAXS geometry, the sampling is, to a very good approximation, orthogonal, but not unitary. Below we derive the full expressions for the reciprocal space sampling vectors in this experimental setup.

Define a coordinate system as follows: Let the incident X-ray beam coincide with the $\mbf{\hat z}$ direction and the sample be at the origin. Denoting the sample-to-detector distance as $D$, then in our experimental geometry the detector lies in the plane $z=D$. Define the polar angle $\alpha$ as the angle with respect to the $z$-axis and $\beta$ as the azimuthal angle in the $xy$-plane (see Fig. \ref{fig:FigS2_sampling}).

\begin{figure}
    \center
    \includegraphics[width=\columnwidth]{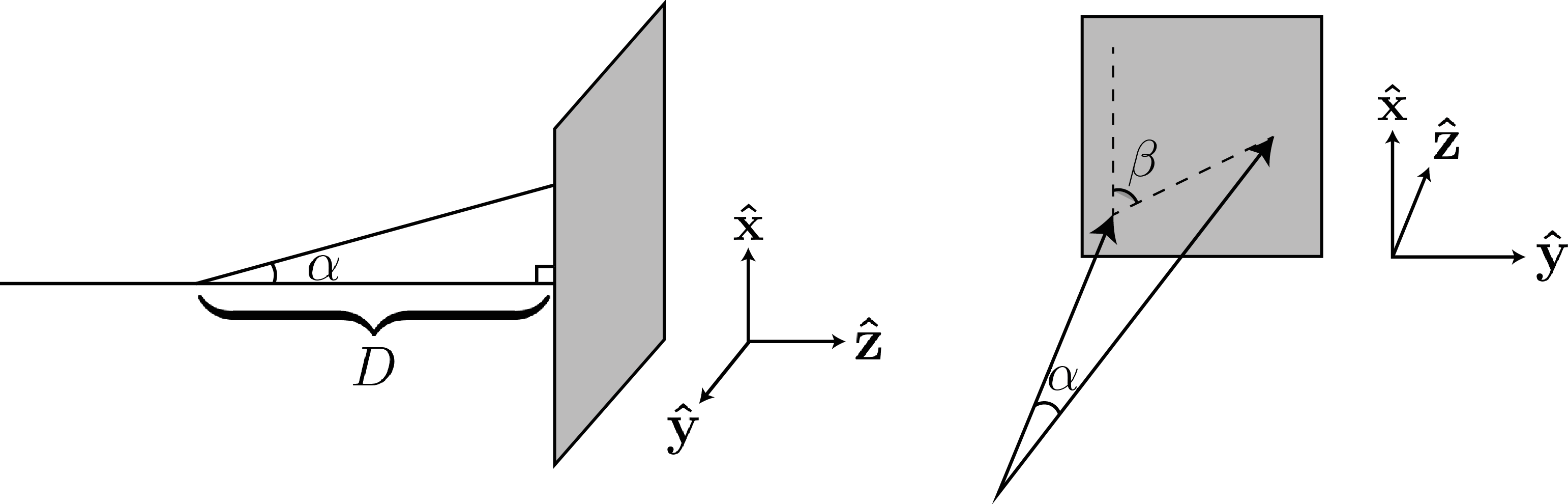}
    \caption{Coordinate system for experimental geometry.}
    \label{fig:FigS2_sampling}
\end{figure}

Let a point $(x,y,D)$ on the detector be assigned to a wavevector via the map $\mbf k:\mathbb R^3\to \mathbb R^3$ given by
\begin{align*}
    (x,y,D) &\xrightarrow{\mbf k} \frac{k}{\sqrt{x^2+y^2+D^2}}\left(x, y, D\right)\\
    &=k(\sin\alpha\cos\beta, \sin\alpha\sin\beta, \cos\alpha)
\end{align*}
where $k$ is a constant equal to $2\pi$ over the radiation wavelength. In particular, $\mbf k(x,y,D)$ is the wavevector parallel to $(x,y,D)$ with magnitude $k$. To compute the reciprocal space step size along the dimensions in the plane of the detector, we calculate:
\begin{align*}
    \frac{\p\mbf q}{\p x}=\frac{\p \mbf k}{\p x} = \frac{\p\mbf k}{\p \alpha}\frac{\p\alpha}{\p x} + \frac{\p\mbf k}{\p\beta}\frac{\p\beta}{\p x}.
\end{align*}

Now $\cos\alpha=\frac{D}{\sqrt{x^2+y^2+D^2}}$; differentiating both sides with respect to $x$ gives
\begin{align*}
    \sin\alpha\frac{\p\alpha}{\p x} = \frac{Dx}{(x^2+y^2+D^2)^{3/2}} = \frac{1}{D}\sin\alpha\cos\beta\cos^2\alpha.
\end{align*}
Similarly, since $\sin\beta=\frac{y}{\sqrt{x^2+y^2}}$, then
\begin{align*}
    \cos\beta\frac{\p\beta}{\p x} = -\frac{xy}{(x^2+y^2)^{3/2}} = -\frac{1}{D}\cos\beta\sin\beta\frac{\cos\alpha}{\sin\alpha}.
\end{align*}
Therefore
\begin{align*}
    \frac{\p\mbf k}{\p x} &= k(\cos\alpha\cos\beta,\cos\alpha\sin\beta,-\sin\alpha)\frac{1}{D}\cos\beta\cos^2\alpha \\
    &-k(-\sin\alpha\sin\beta, \sin\alpha\cos\beta, 0)\frac{1}{D}\sin\beta\frac{\cos\alpha}{\sin\alpha} \\
    &= \frac{k}{D}(\cos^3\alpha\cos^2\beta+\cos\alpha\sin^2\beta,\\
    &\hspace{0.92cm}\cos^3\alpha\sin\beta\cos\beta-\cos\alpha\cos\beta\sin\beta,\\
    &\hspace{0.55cm}-\cos^2\alpha\sin\alpha\cos\beta).
\end{align*}
This is the exact expression. If we take $\alpha$ to be small (as is the case for our SAXS geometry), then keeping all terms up to linear order in $\alpha$,
\begin{align}
    \frac{\p\mbf k}{\p x} &\approx \frac{k}{D}(\cos^2\beta+\sin^2\beta,\sin\beta\cos\beta-\cos\beta\sin\beta, -\alpha\cos\beta) \nonumber\\
    &= \frac{k}{D}(1, 0, -\alpha\cos\beta).
\end{align}

Similarly, one can show
\begin{align}
    \frac{\p\mbf k}{\p y} &= \frac{k}{D}(\cos^3\alpha\cos\beta\sin\beta-\cos\alpha\cos \beta\sin\beta, \nonumber\\
    &\hspace{0.92cm}\cos^3\alpha\sin^2\beta+\cos\alpha\cos^2\beta, \nonumber\\
    &\hspace{0.55cm}-\cos^2\alpha\sin\alpha\sin\beta) \nonumber \\
    &\approx \frac{k}{D}(0,1,-\alpha\sin\beta).
\end{align}
Multiplying by the pixel size $p_x,p_y$ gives the reciprocal space sampling step $\Delta\mbf q_x$, $\Delta\mbf q_x$ along $\mbf{\hat x}$, $\mbf{\hat y}$ respectively.

We now want to calculate the reciprocal space sampling vector that results from the rocking the crystal. In particular, one can show that a rotation of the sample corresponds to an identical rotation of the diffraction pattern. For our experiment, we rotate about the $\mbf{\hat y}$ axis; under a rotation by an angle $\phi$, a point $(q_x,q_y,q_z)$ in reciprocal space gets sent to (under a map we'll call $R_\phi$)
\begin{align*}
    \begin{pmatrix}
        q_x \\ q_y \\ q_z
    \end{pmatrix} \xrightarrow{R_\phi} \begin{pmatrix}
        \cos\phi & 0 & \sin\phi \\
        0 & 1 & 0 \\
        -\sin\phi & 0 & \cos\phi
    \end{pmatrix}\begin{pmatrix}
        q_x \\ q_y \\ q_z
    \end{pmatrix}.
\end{align*}
Then the reciprocal space step $\Delta\mathbf q_\phi$ corresponding to rocking by a small angle $\Delta\phi$ is given by
\begin{align*}
    \Delta\mathbf q_\phi &= \mbf q-R_{\Delta\phi}(\mbf q) \\
    &= (q_x(1-\cos\Delta\phi)-q_z\sin\Delta\phi, \\
    &\hspace{0.62cm}0, \\
    &\hspace{0.62cm}q_x\sin\Delta\phi+q_z(1-\cos\Delta\phi)).
\end{align*}
Rewriting $\mbf q=k(\sin\alpha\cos\beta,\sin\alpha\sin\beta,\cos\alpha-1)$, we have
\begin{align}
    \Delta\mathbf q_\phi
    &= (\sin\alpha\cos\beta(1-\cos\Delta\phi)-(\cos\alpha-1)\sin\Delta\phi)\mbf{\hat x} \nonumber\\
    &+ (\sin\alpha\cos\beta\sin\Delta\phi+(\cos\alpha-1)(1-\cos\Delta\phi))\mbf{\hat z} \nonumber\\
    &\approx k\Delta\phi(0, 0, \alpha\cos\beta),
\end{align}
where again in the last approximation we keep terms only up to linear order in $\alpha$ and $\Delta\phi$.

It is worth noting that the reciprocal space steps, as we've computed them, are all dependent on where in reciprocal space one is. However, since the Bragg peak is fairly small in comparison to its distance from the direct beam and from the sample, we will treat $\alpha$ and $\beta$ as constant, and use their values at the peak to calculate reciprocal space steps that are approximately valid everywhere in the region of reciprocal space used for reconstruction.

For the experimental dataset presented in the main text, we have  $\alpha\approx0.0193, \beta \approx-0.541$. Then $|\Delta\mbf q_x|=|\Delta\mbf q_y|\approx776300\T{ m}^{-1}$, and  $|\Delta\mbf q_\phi|\approx 270200\T{ m}^{-1}$. Moreover, we can check that the reciprocal space sampling vectors are nearly orthogonal:
\begin{align*}
    \cos^{-1}\left(\frac{\lb \Delta\mbf q_x,\Delta\mbf q_y\rb}{|\Delta\mbf q_x||\Delta\mbf q_y|}\right) &\approx 90.0^\circ, \\
    \cos^{-1}\left(\frac{\lb \Delta\mbf q_y,\Delta\mbf q_\phi\rb}{|\Delta\mbf q_y||\Delta\mbf q_\phi|}\right) &\approx 90.6^\circ, \\
    \cos^{-1}\left(\frac{\lb \Delta\mbf q_\phi,\Delta\mbf q_x\rb}{|\Delta\mbf q_\phi||\Delta\mbf q_x|}\right) &\approx 90.3^\circ.
\end{align*}

\newpage
\subsection*{CDI analysis and visualization methods}The Python package Phaser was used to carry out error reduction (ER), hybrid input-output (HIO), and shrinkwrap algorithms \cite{siddharth_maddali_phaser_2020}. All reconstructions were done on the Agave and Sol computing clusters at Arizona State University. The reconstruction in Fig. \ref{fig:Fig2_CDI}b was produced using an initial support that is a sphere of radius 1.5 $\mu$m. We used the following algorithm to obtain reconstructions: 3x[500 iterations ER, 200 iterations HIO], 500 iterations ER, 10x[Shrinkwrap/intersect, 5x[500 iterations ER, 200 iterations HIO], 500 iterations ER], 10x[500 iterations ER, 200 iterations HIO], 2000 iterations ER. Here, the Shrinkwrap/intersect step consists of four steps: (1) applying a Gaussian filter of standard deviation 3 voxels to the real space object, (2) assigning only those voxels above a threshold of 0.025 times the maximum voxel value to a preliminary support region, (3) computing the diameter (along one axis) of the pre-existing support, and then (4) intersecting the preliminary support region from step (2) with a sphere of diameter 1.2 times the pre-existing support diameter computed in (3), and positioned, for the $i$th axis, at the coordinate corresponding to the largest cross-section along axis $i$ of the pre-existing support. This intersection is taken to be the updated support. A $\beta$ value of 0.99 was used for the HIO steps. Finally, in visualizing the reconstruction in Fig. \ref{fig:Fig2_CDI}b we use nearest-neighbor interpolation to correct for the non-unitary sampling in real space. The sampling vectors are sufficiently close to orthogonal that we do not correct for non-orthogonality.

For the reconstructions featured in Fig. \ref{fig:Fig2_CDI}d we use the following algorithm: 10x[500 iterations ER, 200 iterations HIO, 500 iterations ER, 500 iterations HIO], 200 iterations ER. Each reconstruction support is a sphere of fixed diameter throughout the entire reconstruction process. Five reconstructions were completed at each support diameter, and their line cuts plotted individually in Fig. \ref{fig:Fig2_CDI}d. Each line cut has an in-plane width of three pixels, and is averaged over five angular (out-of-plane) steps. Each reconstructed intensity pattern was aligned with the data prior to taking a line cut.

The three-dimensional density visualization in Fig. \ref{fig:Fig2_CDI}b is described in the main text as a “surface visualization,” by which we mean the following: We first take the magnitude of the complex reconstructed density. Then we divide the density range into bins that are equally spaced on a logarithmic scale. For each bin, we create a surface (with Paraview’s surface representation) using only the voxels with a density range falling into that bin, and with opacity scaled by the density values.

\newpage
\subsection*{Additional CDI analysis} Fig. \ref{fig:FigS3_135RSL} presents real space line-cuts (each of which has a width of five pixels) through the 3D reconstructed density shown in Figure 2b. One can see in clearer fashion (than with a three-dimensional visualization) the decay in density away from the center of the object.

\begin{figure}
    \center
    \includegraphics[width=\columnwidth]{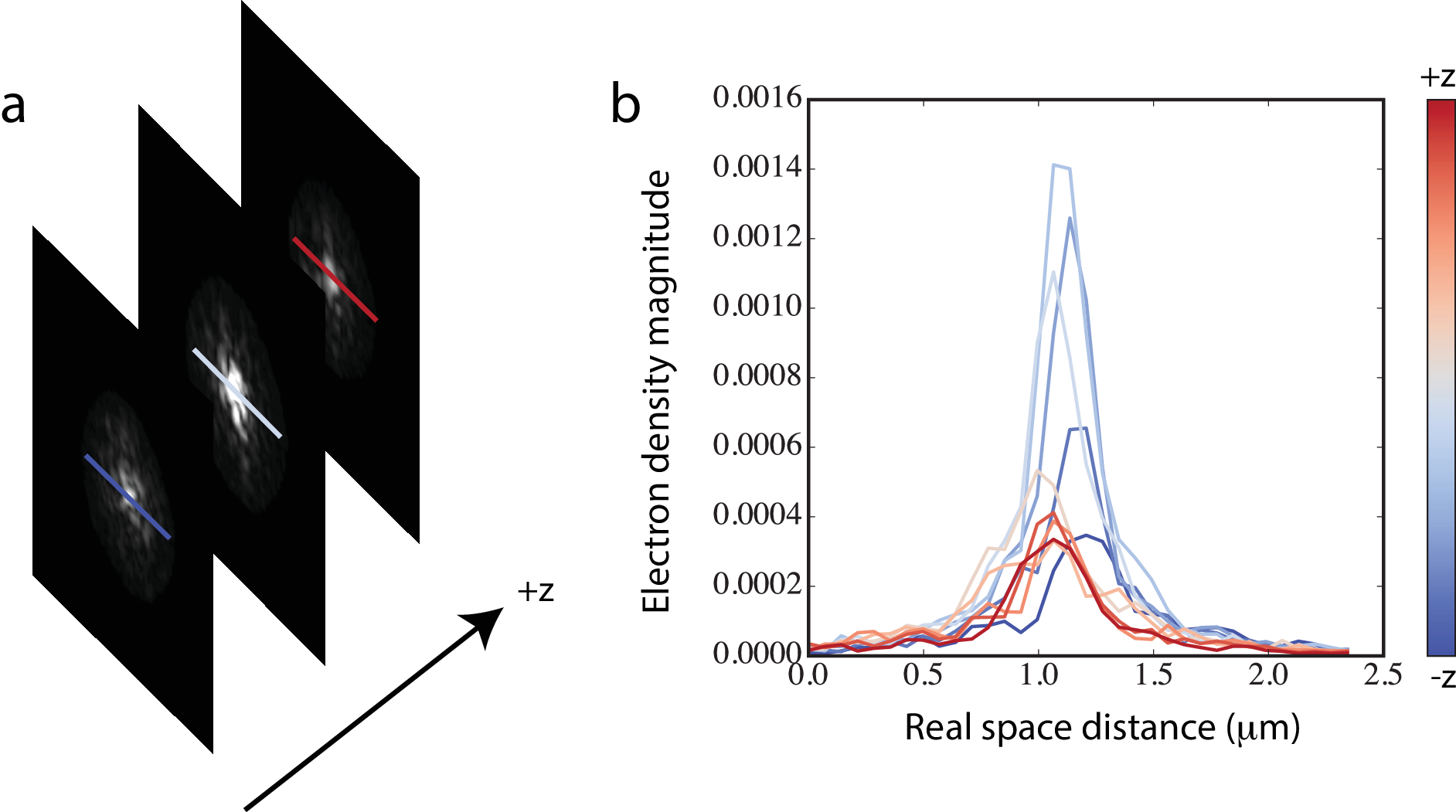}
    \caption{Visualization of real space CDI reconstruction. (a) Several planar cross-sections of the density in Fig. \ref{fig:Fig2_CDI}b, with a horizontal line cut indicated through the center of each. (b) A plot of the resulting line cuts over eight such cross-sections, showing density as a function of distance across the line cut. The color of each trace indicates the cross-section to which its line-cut corresponds.}
    \label{fig:FigS3_135RSL}
\end{figure}

Fig. \ref{fig:FigS4_116cdi} presents results from CDI analysis on another SL in the same capillary as that of the main text. Here we use a 92 $\times$ 92 pixel ROI for reconstruction, again with a rocking curve consisting of 128 angular steps covering two degrees in total. The reconstruction in Fig. \ref{fig:FigS4_116cdi}b uses an initial spherical support of radius 1.4 $\mu$m and is produced via the following algorithm: 3x[500 iterations ER, 200 iterations HIO], 500 iterations ER, 10x[Shrinkwrap/intersect, 10x[500 iterations ER, 200 iterations HIO], 500 iterations ER], 10x[500 iterations ER, 200 iterations HIO], 2000 iterations ER. The Shrinkwrap/intersect step proceeded as described in the \textbf{CDI analysis methods} section above, except with a threshold factor of 0.075 instead of 0.025. The reconstruction line cuts featured in Fig. \ref{fig:FigS4_116cdi}d employ the same methodology of comparing support sizes as those of Fig. \ref{fig:Fig2_CDI}d.

\begin{figure}
    \center
    \includegraphics[width=\columnwidth]{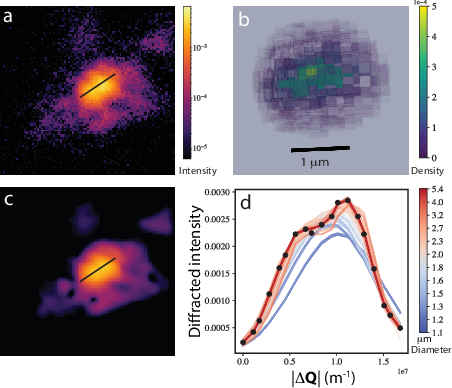}
    \caption{Coherent diffractive imaging of a secondary SL. (a) Detector image showing a 2D cross-section through the middle of the 3D intensity dataset, with a line cut of interest marked in black. (b) Surface visualizations of the reconstructed electron density, with scaled opacity. Yellow regions correspond to higher density and purple to lower density. Scale bar in black, indicating a distance of 1 $\mu$m. (c) Corresponding 2D cross-section of the reconstructed diffraction pattern. (d) Line cuts of the measured and reconstructed diffraction patterns. Black dots are the measured intensity as a function of distance in reciprocal space $\Delta $ along the line cut in (a). Similarly, colored traces correspond to line cuts taken from reconstructed intensities, with color indicating the reconstruction support diameter.}
    \label{fig:FigS4_116cdi}
\end{figure}

Fig. \ref{fig:FigS5_116RSL} plots real-space line cuts through the reconstructed object in Fig. \ref{fig:FigS4_116cdi}b, akin to those of Fig. \ref{fig:FigS3_135RSL} for the SL in Fig. \ref{fig:Fig2_CDI} of the main text.

\begin{figure}
    \center
    \includegraphics[width=1\columnwidth]{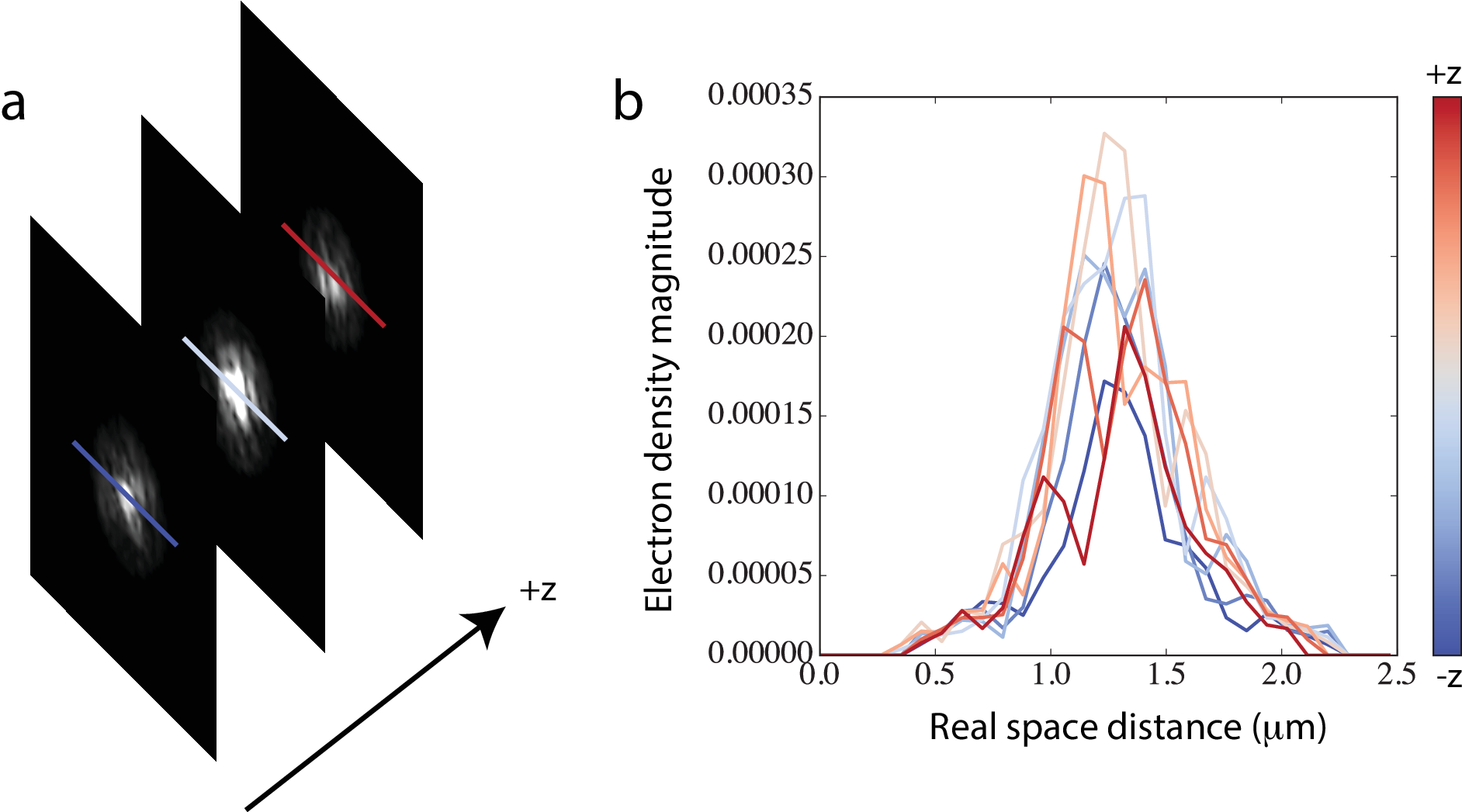}
    \caption{Visualization of real space CDI reconstruction for a secondary SL. (a) Several planar cross-sections of the density in Fig. \ref{fig:Fig2_CDI}b, with a horizontal line cut indicated through the center of each. (b) A plot of the resulting line cuts over eight such cross-sections, showing density magnitude as a function of distance across the line cut. The color of each trace indicates the cross-section to which its line-cut corresponds.}
    \label{fig:FigS5_116RSL}
\end{figure}

\newpage
\subsection*{Estimates of wavefront distortion due to quartz wool}

Wavefront distortions on the X-ray beam due to an inhomogeneous medium can occur either before or after scattering from the SL. We argue that the wavefront distortions before diffraction are less important, as the SL acts essentially as a spatial filter on the focused beam, where only the wavefront in the $\sim\mu$m scale cross section of the SL will have effects on the final diffraction pattern.  Given that no structures in the beam (quartz wool, or the capillary itself) are on the length scale of the reconstructed SL structure, it is unlikely that any wavefront errors would produce structure in the reconstruction.

To estimate the wavefront distortion of the quartz wool, we need to estimate (1) the number of fibers in the diffracted beam, (2) the size of the fibers, and (3) the Rayleigh range of the diffracted beam. The Rayleigh range of the diffracted X-ray beam from the SL (111) peak sets the thickness through which we can treat wavefront distortions as multiplicative near-field errors.  The Rayleigh range $Z_r$ is defined by $Z_{r} = \pi \frac{w_0^2}{\lambda}$, where $w_0$, in this case, is the half-width of the smallest structure in the SL ($w_0 \approx 200$ nm based on our reconstruction). This sets a $Z_r = 1.25$ mm, longer than internal capillary thickness. Therefore, we can safely assume that any inhomogeneities in the beam occur in the near-field regime, where the accumulated phase is simply added to the SL phase.

We estimate that the fibers have approximately a $30\%$ fill fraction, based on the volume of fluid required to fill a packed capillary. The quartz wool consists, on average, of cylinders on the order of 10 $\mu$m thick. Therefore, one would expect $0.3 \times 1000/10 = 30$ fibers, on average, through the entire length of the capillary.  Assuming the SL lies in the middle of the capillary, there would be 15 fibers in the diffracted beam.  Since the radius of curvature of the fibers is much larger than the size of the diffracted beam in the near field ($10 \T{ }\mu\T{m} \gg 500 \T{ nm}$) we assume that unless the diffracted beam strikes the very edge of the fiber, it acts like a diverging lens with a focal length of $f = -2.0$ meters. The focal length is calculated using the known real part of the refractive index of quartz ($\T{Re}(n) = 1 - 3.98 \times 10^{-6}$) and hydrazine ($\T{Re}(n) = 1 - 1.54 \times 10^{-6}$) and the lensmaker's equation. Such a lens placed 0.5 mm from the object would form a virtual image 0.12 $\mu$m from the original object position, a negligible displacement even on the scale of the object. Therefore, we can neglect any astigmatism introduced through transmission through the large radius of curvature fibers.

Next we must consider the possibility that the diffracted beam strikes an edge of the fiber, rather than the center. The probability that a 400 nm full-width wide beam strikes the edge of a $10 \T{ }\mu$m wide fiber is $P = 400/10,000 = 0.04$. With, on average, 15 chances to occur, one for each fiber in the beam, the total probability of striking a single edge where only part of the diffracted beam is in glass is $45\%$. Therefore, while there is a reasonable probability that a single wavefront distortion event took place, the distortion from a single transmission is small.

If such an event were to occur, the phase shift $\Delta\phi$ would be defined by

\begin{equation}
    \Delta \phi  = \frac{2 \pi }{\lambda} d (n_1 - n_2),
\end{equation}

where $n_1$ and $n_2$ are the indices in hydrazine and quartz, and $d$ is the path length inside the quartz. Given that the beam that strikes an edge is 200 nm in the quartz at most, the maximum path length in quartz is 2.8 $\mu$m (the length of the chord). This gives a total phase shift of  $\Delta\phi = 0.42$ radians, which is less than $\lambda/10$, an acceptable wavefront error for reconstruction. This analysis indicates that the reconstructions of the SL structure are unlikely to be dominated by wavefront errors due to the quartz wool phase.

\newpage
\section*{XPCS Methods}
\subsection*{XPCS data collection and autocorrelation computation}

XPCS data collection proceeded by collecting data at a repetition rate of 120 Hz; the results in Fig. \ref{fig:Fig3_XPCS_oscillations} and Fig. \ref{fig:Fig4_XPCS_avalanches} are from a dataset consisting of 17,329 shots (144 seconds). All one-time and two-time correlations were calculated using the Python package scikit-beam \cite{abeykoon_software_2016}. For an intensity distribution $I(\textbf Q,t)$, scikit-beam calculates the temporal autocorrelation over a given region of interest to be
\begin{equation}\label{g2 equation}
    g^{(2)}(\tau) = \frac{\lb I(\mbf Q,t)I(\mbf Q,t+\tau)\rb_{\mbf Q,t}}{\lb I(\mbf Q,t)\rb_{\mbf Q,t}^2},
\end{equation}
where the brackets $\lb\rb_{\textbf Q,t}$ denote an average over both time and $\textbf Q$. Moreover, scikit-beam employs the so-called symmetric normalization scheme: for a sequence $(I_t)_{t=1}^N$ of intensity values, the autocorrelation computed with symmetric normalization is given by
\begin{equation}\label{symmetric_norm}
    g^{(2)}(k) = \frac{\frac{1}{N-k}\sum_{t=1}^{N-k} I_tI_{t+k}}{\left(\frac{1}{N-k}\sum_{t=1}^{N-k}I_t\right)\left(\frac{1}{N-k}\sum_{t=k+1}^NI_t\right)}.
\end{equation}

\subsection*{SL annealing XPCS analysis methods}

We divide the detector into ring-shaped regions of width 2.5 pixels, centered at a given point the peak, and further divide each ring into longitudinal and transverse sections. To compute the location of $\mbf Q=0$ (which is blocked by a beam stop), we fit the (111) diffraction ring to a circle, and take the direct beam to be at the center. The longitudinal direction is then given by the vector along the detector from the direct beam to the ring center, and the transverse direction orthogonal to the longitudinal vector. We then calculate an autocorrelation for each ROI and fit it to a stretched exponential form $g^{(2)}(\tau) = k + \beta\exp(-(\Gamma'\tau)^\alpha)$, choosing fit ranges for each ROI to capture any decay that exists while avoiding the initial oscillatory behavior and any late-time artifacts. Fig. \ref{fig:FigS6_radialfits} and Fig. \ref{fig:FigS7_radialfits} show example autocorrelations and fits for the longitudinal and transverse directions, respectively.

\begin{figure*}
    \includegraphics[width=\textwidth]{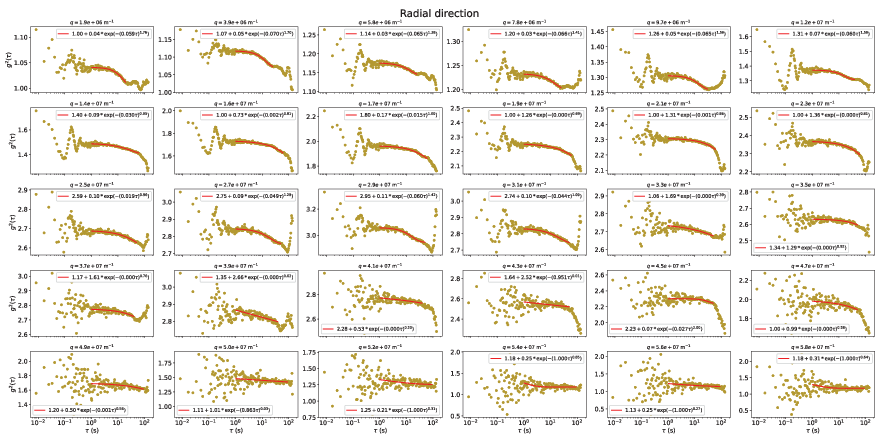}
    \caption{Autocorrelations computed along the longitudinal direction (gold points) as a function of $q$ with corresponding fits (red curves).}
    \label{fig:FigS6_radialfits}
\end{figure*}

\begin{figure*}
    \includegraphics[width=\textwidth]{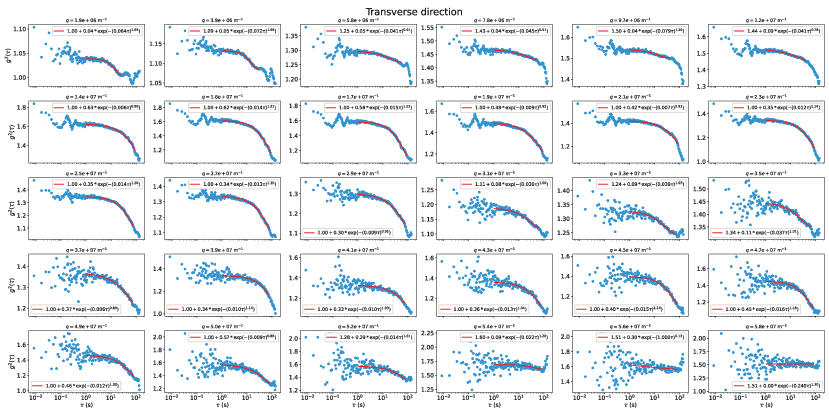}
    \caption{Autocorrelations computed along the transverse direction (blue points) as a function of $q$ with corresponding fits (red curves).}
    \label{fig:FigS7_radialfits}
\end{figure*}

From these fits we obtain a dispersion in the decorrelation rate and stretching exponent (Fig. \ref{fig:FigS8_singlecenterdispersion}). We calculate $\Gamma=0.5^{1/\alpha}\Gamma^{\prime}.$

\begin{figure}[H]
    \centering
    \includegraphics[width=\columnwidth]{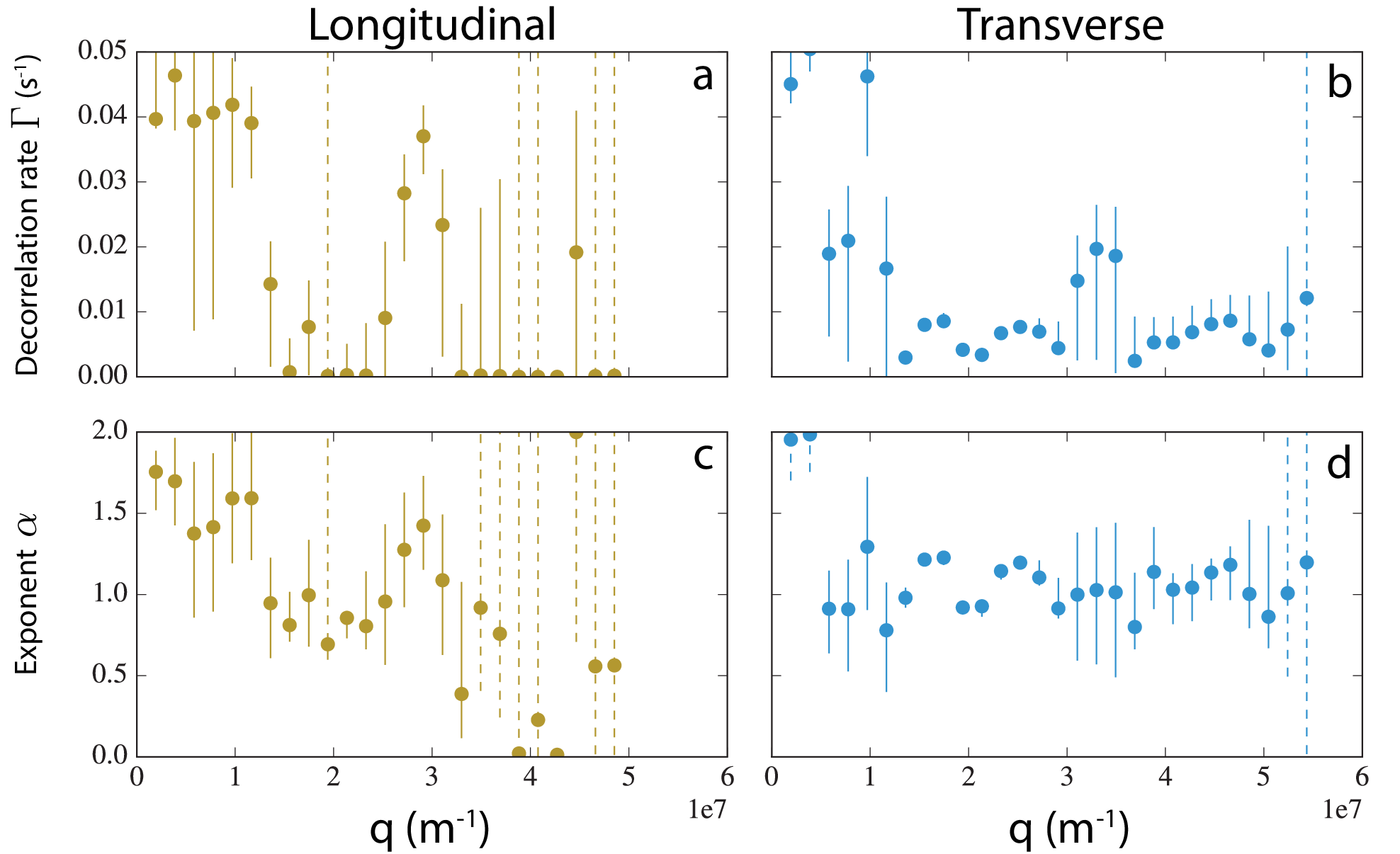}
    \caption{Dispersion results for fits around a single choice of $q=0$. (a, b) Decorrelation rate $\Gamma$ as a function of $q$ in the longitudinal and transverse directions, respectively. Confidence intervals for each fit parameter are shown as error bars; dashed error bars indicate that the upper bound on the confidence interval is infinity. (c, d) Stretching exponent $\alpha$ as a function of $q$ in the longitudinal and transverse directions, respectively.}
    \label{fig:FigS8_singlecenterdispersion}
\end{figure}

Figs. \ref{fig:FigS6_radialfits}-\ref{fig:FigS8_singlecenterdispersion} are all calculated for a single choice of center ($q=0$). As described in the main text, we then repeat this process for 81 different choices of center, once for each pixel within 5 pixels of the intensity center of mass of the time-integrated intensity pattern. We use the same fit ranges for each choice of center. We then remove outliers from the fit results (points at least 1.5 times the interquartile range larger/smaller than the upper/lower quartile) and average the remaining parameters together to get the final results.

Fitting to a simple exponential decay (setting $\alpha=1$) rather than a stretched exponential yields the dispersion featured in Fig. \ref{fig:FigS9_fixedfits}. In particular, the increasing dispersion in $\Gamma$ along the transverse direction for large values of $q$ persists when using a fixed stretching exponent, and such dispersion is not seen in the longitudinal direction.

\begin{figure}[H]
    \centering
    \includegraphics[width=5cm]{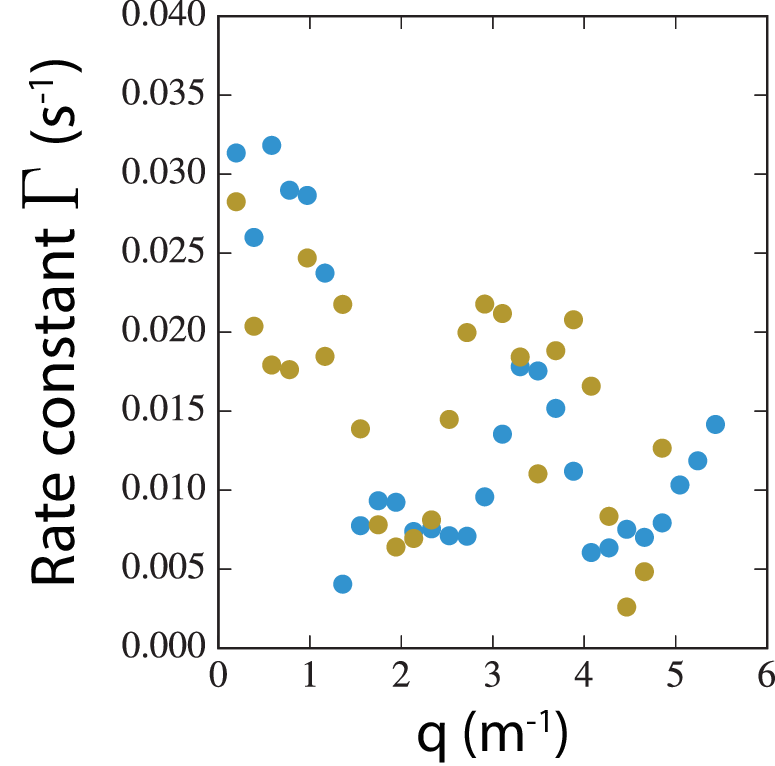}
    \caption{Dispersion of decorrelation rate $\Gamma$ vs. $q$ for the longitudinal (gold) and transverse (blue) directions with a fixed stretching exponent of 1.}
    \label{fig:FigS9_fixedfits}
\end{figure}

\clearpage
\section*{Appendix F: Resolution of features in the diffraction pattern and upper bounds on SL angular motion}

As discussed in the main text, the superlattices are not completely immobilized in the quartz wool. Challenges to extracting the SL motion from the data directly include: the time-resolved data is a 2D cross-section of the 3D diffraction pattern; rotations do not easily map onto a rectilinearly-sampled dataset, and it is not straightforward to determine intensity fluctuations from an intensity autocorrelation. We can, however, attempt to place limits on the extent of the motion using the resolution of features in the measured static 3D diffraction pattern.

In particular, the three-dimensional diffraction data is indexed by two in-plane pixel dimensions (forming detector images), and one out-of-plane dimension (which corresponds to rotations of the sample). We plot various features of this 3D dataset below.

Fig. \ref{fig:FigS10_rockingcurves} plots rocking curves (intensity as a function of rotation angle) of three different pixels on the detector. The peaks in (b) and (d) each have a FWHM of less than six angular steps. Moreover, the peaks are not perfectly symmetric; in each case one side features a sharper drop than the other and the curve drops to less than half of its maximum value in only two angular steps. Note that such a width of five angular steps corresponds to an interval of $0.078^\circ$ in the rotation angle $\phi$. Additionally, the reciprocal space distance between the two peaks in (c) is approximately $|\Delta \mbf Q|=2.2 \T{ }\mu \T{m}^{-1}$, such that $2\pi/|\Delta \mbf Q|\approx 2.9\T{ }\mu$m.

\begin{figure}
    \centering
    \includegraphics[width=\columnwidth]{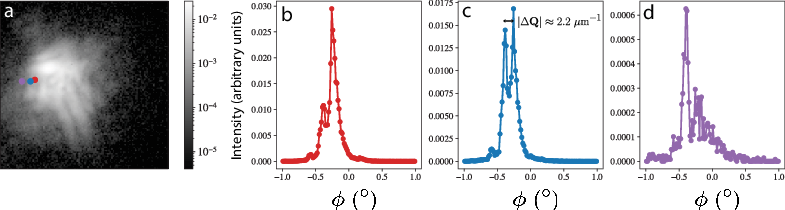}
    \caption{Out-of-plane features of the 3D rocking curve dataset. (a) 2D cross section (detector image) of SL (111) Bragg peak intensity analyzed in main text. (b-d) Intensities of individual pixels marked in (a) as a function of angular step $\phi$.}
    \label{fig:FigS10_rockingcurves}
\end{figure}

Fig. \ref{fig:FigS11_inplanelinecuts} plots in-plane line cuts from two different detector images (cross-sections of the same 3D diffraction pattern investigated in Fig. \ref{fig:FigS10_rockingcurves}). Each features peaks with FWHM around five pixels or less (note that five pixels corresponds to a reciprocal space distance of  $3.9 \T{ }\mu\T{m}^{-1}$). Furthermore, a rotation of the SL of only approximately 0.25$^\circ$ about the direct beam would shift a
feature in the diffraction pattern by five pixels in the transverse direction, since the diffraction pattern is about 1250 pixels from the direct beam.

\begin{figure}
    \centering
    \includegraphics[width=\columnwidth]{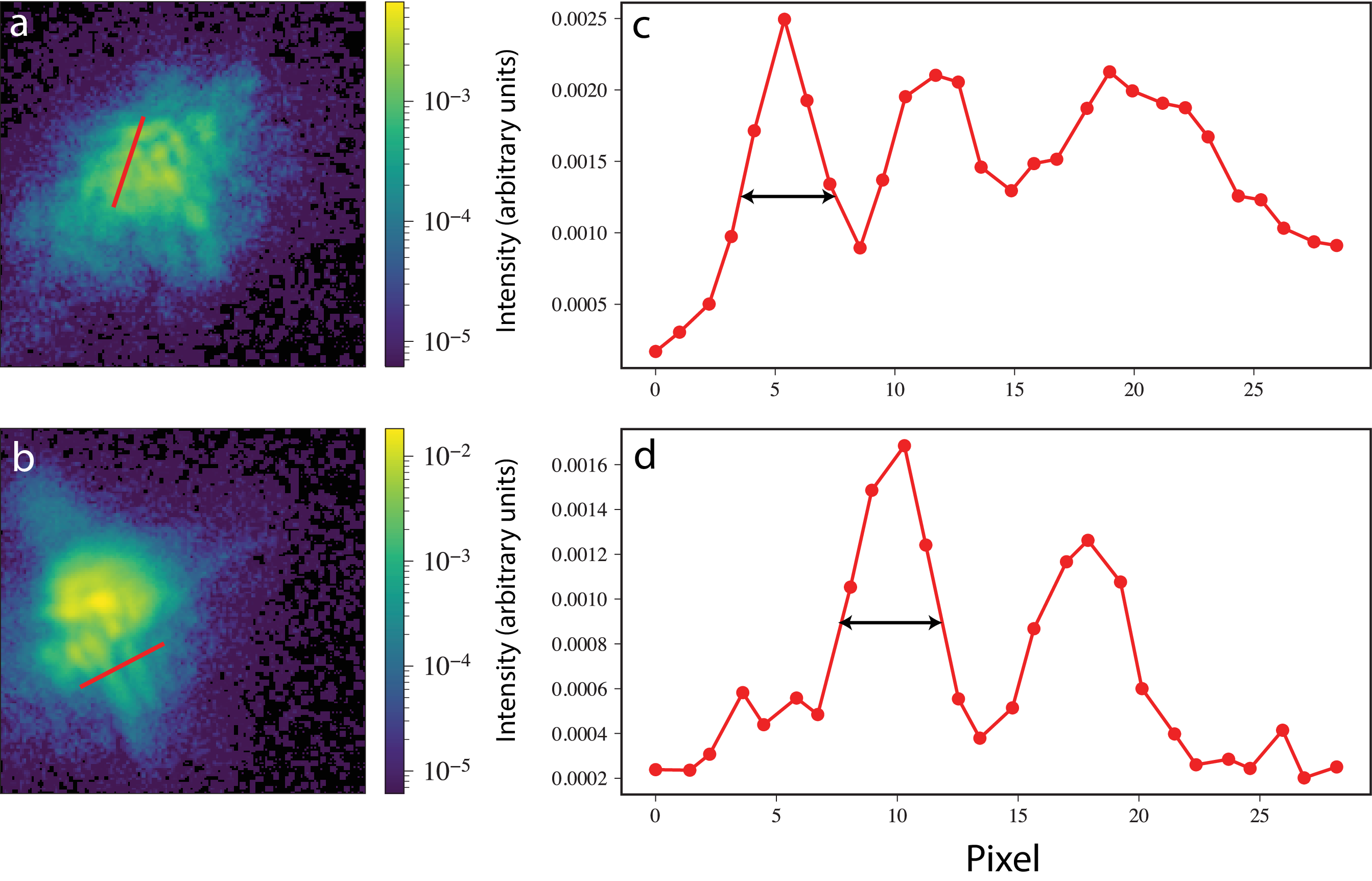}
    \caption{In-plane features of the 3D rocking curve dataset. (a,b) Detector images. (c,d) Line cuts of interest (in-plane intensity as a function of pixel) marked in red on images in (a) and (b), respectively.}
    \label{fig:FigS11_inplanelinecuts}
\end{figure}

Since the data acquisition period (2 s) for each rocking angle is roughly an order of magnitude larger than the autocorrelation oscillation period, these fringe widths serve as upper bounds on the extent to which our intensity dataset may be blurred by SL librations. That is, we expect the SL rotation about the rocking axis to be no more than about 0.08$^{\circ}$ (a maximum motion of a 4 nm arc length for a 3 $\mu$m structure) and the rotation about the direct beam to be no more than about 0.25$^{\circ}$ (a maximum motion of a 13 nm arc length for a 3 $\mu$m structure). We cannot easily quantify rotations about a third axis orthogonal to these first two, but since there is no reason to expect that the superstructure would preferentially oscillate about such an axis, we assume a similar value for it as well.

\clearpage
\section*{Appendix G: Additional XPCS analysis}

We observe a square-like pattern in the two-time correlation function in another experimental dataset (Fig. \ref{fig:FigS12_ttcf93}).

\begin{figure}
    \centering
    \includegraphics[width=\columnwidth]{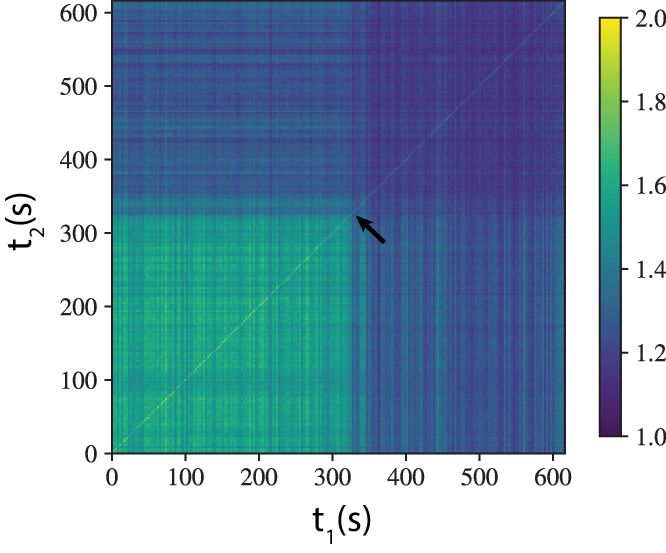}
    \caption{Two-time correlation function from a second experimental dataset, computed over a ring of inner and outer radius $q=1.0\cdot10^7 \T{ m}^{-1}$ and $q=1.2\cdot10^7 \T{ m}^{-1}$, respectively.}
    \label{fig:FigS12_ttcf93}
\end{figure}

Furthermore, we observe oscillatory behavior in the autocorrelation functions from other datasets, with the strongest frequencies consistently occurring in the range 4 Hz to 7 Hz. Fig. \ref{fig:FigS13_XPCS_Oscillations_115} repeats the analysis of the main text for another SL. Here autocorrelations were calculated over 11 $\times$ 11 pixel ROIs to produce the frequency map in Fig. \ref{fig:FigS13_XPCS_Oscillations_115}d.

\begin{figure}
    \centering
    \includegraphics[width=\columnwidth]{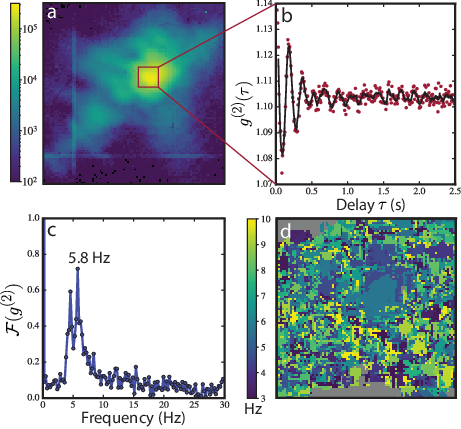}
    \caption{Short-time oscillatory behavior in the SL (111) intensity autocorrelation function for a secondary SL. (a) Cropped detector image featuring the (111) Bragg peak. (b) Intensity autocorrelation calculated over the red ROI marked in (a). (c) Spectral density of the signal in (b). (d) Frequency map of the strongest oscillatory component in the autocorrelation spectral density.
}
    \label{fig:FigS13_XPCS_Oscillations_115}
\end{figure}

In this third dataset (which consisted of 23978 shots, or about 200 seconds worth of data), we do not observe any decay in the intensity autocorrelation functions. Fig. \ref{fig:FigS14_115g2s} displays autocorrelations as a function of the radius of the annular ROIs over which they were calculated.
\begin{figure}
    \centering
    \setcounter{figure}{19}
    \includegraphics[width=\columnwidth]
    {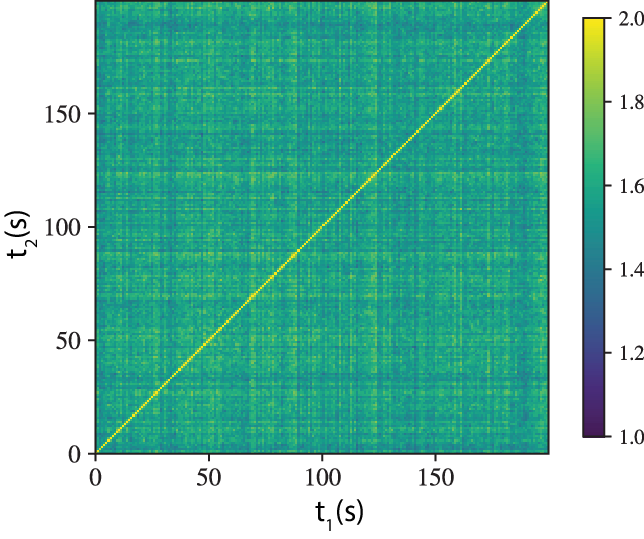}
    \caption{Two-time correlation function from the same dataset as featured in Fig. \ref{fig:FigS13_XPCS_Oscillations_115} and \ref{fig:FigS14_115g2s}.}
    \label{fig:FigS15_115ttcf}
\end{figure}

\begin{figure*}
    \centering
    \setcounter{figure}{18}
    \includegraphics[width=\textwidth]{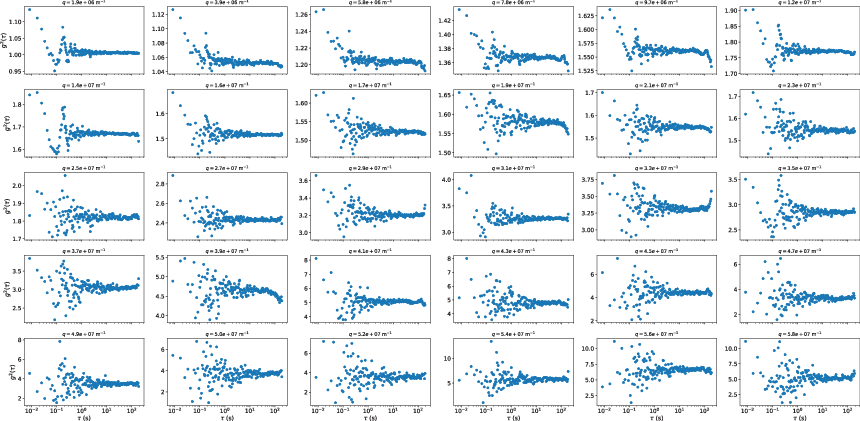}
    \caption{Autocorrelations computed as a function of $q$ for the additional SL dataset from Fig. \ref{fig:FigS13_XPCS_Oscillations_115}.}
    \label{fig:FigS14_115g2s}
\end{figure*}

Fig. \ref{fig:FigS15_115ttcf} shows a two-time correlation function from this same dataset computed over a ring of inner and outer radius $q=1.7\cdot10^7 \T{ m}^{-1}$ and $q=1.9\cdot10^7 \T{ m}^{-1}$. In particular, we note that, in contrast to the two-time correlation shown in Fig. \ref{fig:Fig4_XPCS_avalanches}f, this correlation function does not present sharp discontinuities. The absence of such jumps, accompanied by the lack of decay in the one-time autocorrelations, suggests that the wavevector-dependent time scale observed in Fig. \ref{fig:Fig4_XPCS_avalanches}d is the product of the intensity shifts that show up as discontinuities in the two-time correlation plot in Fig. \ref{fig:Fig4_XPCS_avalanches}f.


\section*{Appendix H: Simulation of hypothetical single crystal and polycrystalline \textit{fcc} lattices}

We arrange 19619 atoms into an \textit{fcc} lattice. To produce the polycrystal, we randomly select 20 dividing planes that are either (111), (110), or (100) and, for each one, rotate one side of the crystal by a random angle, and shift the rotated domain a particle radius away perpendicular to the plane, to prevent particle overlap. We then minimize the energy of the resulting grain boundaries using a pairwise Lennard-Jones potential. From here, we use the NC positions to compute the structure factor and multiply by a hard-sphere form factor to obtain the complex scattering pattern. We then crop this scattering pattern to a cubic region of interest (with side length $3.4\cdot10^8 \T{ m}^{-1}$) containing the Bragg peak, and perform a numerical inverse Fourier transform. Taking the complex magnitude then yields the density plots in Fig. \ref{fig:Fig5_simulations}e and \ref{fig:Fig5_simulations}f.

\newpage
\bibliography{LU89_paper_2023v2.bib}


\end{document}